\documentclass[twocolumn,superscriptaddress,amsmath,amssymb,aps,prb,floatfix]{revtex4-2}
\usepackage{ragged2e}

\usepackage{graphicx}
\usepackage{xcolor}
\usepackage{overpic}
\usepackage{braket}
\usepackage{bm}
\usepackage{comment}
\usepackage[breaklinks=true,colorlinks,citecolor=blue,linkcolor=blue,urlcolor=blue]{hyperref}
\usepackage{soul}
\usepackage{tabularx}  % Required for variable-width columns
\usepackage{booktabs}  % For better-looking tables
\usepackage{makecell}  % For line breaks within cells
\usepackage{tikz} 
\usepackage{tikz-feynman}
\usepackage{circuitikz}
\usepackage{bbm}

\usepackage{cancel}
\usepackage{subcaption}

\newcommand{\parent}[1]{\left(#1\right)}
\newcommand{\commut}[2]{\left[ #1, #2\right]}
\newcommand{\ii}{\mathrm{i}}
\newcommand{\Z}{\mathbb{Z}}
\newcommand{\N}{\mathbb{N}}
\newcommand{\R}{\mathbb{R}}

\renewcommand{\P}{\mathbb{P}}
\newcommand{\1}{\mathbbm{1}}

\begin{document}

%
%title{Effective description of electrons coupled to an off-resonant cavity}
\title{Floquet Theory of lattice electrons coupled to an off-resonant cavity}

\author{Jules Sueiro}
\affiliation{JEIP, UAR 3573 CNRS, Collège de France, PSL Research University, 11 Place Marcelin Berthelot,  F-75321 Paris, France}

\author{Gian Marcello Andolina}
\affiliation{JEIP, UAR 3573 CNRS, Collège de France, PSL Research University, 11 Place Marcelin Berthelot,  F-75321 Paris, France}

\author{Marco Schir{\`o}}
\affiliation{JEIP, UAR 3573 CNRS, Collège de France, PSL Research University, 11 Place Marcelin Berthelot,  F-75321 Paris, France}

\begin{abstract}
We use Floquet theory and the High-Frequency expansion to derive an effective Hamiltonian for electrons coupled to an off resonant cavity mode, either in its vacuum or driven by classical light. For vacuum fields, we show that long-range hopping and cavity-mediated interactions arise as a direct consequence of quantum fluctuations. As an application, this method is applied to the Su-Schrieffer-Heeger (SSH) model. At high light-matter coupling, our results reveal significant deviations from mean-field predictions, with our framework capturing light-matter entanglement through the Floquet micromotion. Furthermore, the cavity-mediated interactions appearing at first order are shown to be crucial to the description of the system at sufficiently strong light-matter coupling for a fixed cavity frequency. Finally, a drive resonant with the cavity is added with the SSH chain displaying dynamical behavior dependent on the cavity parameters.
%We show for vacuum fields the leading effect are long-range hopping and cavity mediated interactions, while the cavity mode remains unpopulated. In the driven case the physics is more reach and a non trivial population of the cavity mode is possible. We apply this formalism to different problems including..
\end{abstract}

%The structure of the Hamiltonian of matter coupled to a single-mode cavity exhibits notable similarities to that of a classically driven system in the Floquet extended space.

%Following this similarity, we use Floquet theory -specifically, the high-frequency expansion- to derive an effective Hamiltonian for tight-binding electrons coupled to an off-resonant cavity through the full Peierls phase. The expansion derived is controlled by the cavity frequency rather than the light-matter coupling, allowing the exploration of the strongly coupled regime. Finally, the Floquet-based method is easily adapted to accommodate the addition of a classical drive resonant with the cavity.

\maketitle

\section{Introduction}\label{sec:intro}

%Light has emerged as the primary way to control the state of an atom. 
In the last decades experimental advances in controlling light-matter interactions have allowed to generate ultra-fast pulses of tunable frequency and intensity and use them to probe and control matter. The development of these so-called pump-probe techniques has thus opened the pathway to extend light-control of quantum states from atoms and molecules~\cite{mukamel1990femto} to quantum materials systems~\cite{giannetti2016ultrafast,nicoletti2016nonlinear}. Using tailored pulses of light has allowed to induce topology \cite{mciver_light-induced_2020} or even drive phase transitions in condensed-matter systems, such as light-induced superconductivity~\cite{fausti_light-induced_2011,mitrano2016possible}. 
Theoretical descriptions of the effect of driving the system have explored the use of Floquet theory and Floquet engineering, as a way to stabilize new phases of matter of out equilibrium~\cite{oka_floquet_2019}.

%rely on Floquet theory which allows to obtain the long-time dynamics of a periodically-driven system.

By confining light in a cavity, a single photon creates an important electric field, such that matter can feel the granularity of the electromagnetic field. For over two decades, Cavity Quantum Electrodynamics (C-QED) has demonstrated the effect of a quantized cavity field on an atom's quantum state \cite{brune_quantum_1996}. Similarly to classical light, one can imagine to use quantum light to control the properties of an extended system \cite{schlawin_cavity_2022}. Theoretical proposals are numerous and range from the cavity-enhancement of superconductivity \cite{curtis_cavity_2019} to the modification of the topology of materials \cite{wang_cavity_2019,ciuti2021cavity,dmytruk_controlling_2022,nguyen2024electron,rokaj2023weakened}. Experimentally, cavities have demonstrated the control of a metal to insulator transition \cite{fausti_cavity-mediated_2023}, the alteration of chemical reaction \cite{garcia-vidal_manipulating_2021} and the modification of transport and the quantum Hall effect in a two-dimensional electron gas \cite{paravicini-bagliani_magneto-transport_2019,appugliese_breakdown_2022,enkner_nature2025}.

Light being a quantum field, even in the absence of photons in the cavity, i.e. the cavity field is in its vacuum state, the electromagnetic field fluctuates. Those fluctuations are at the origin of the effects measured in dark cavities like those mentioned above. Through those fluctuations, the cavity can mediate interactions between the electrons or renormalize the constants of the problem. To effectively describe matter strongly coupled to a cavity, our theories need to incorporate the fluctuations of the field. Different theoretical approaches have been developed in the past few years to tackle the resulting photon-electron problem. On the analytical front many-body techniques such as mean-field theory plus gaussian fluctuations or diagrammatic methods have been used~\cite{mazza_superradiant_2019,andolina_theory_2020,guerci_superradiant_2020,dmytruk_gauge_2021,lenk2022collectivetheoryinteractingsolid,lenk2022dynamical,dmytruk_controlling_2022,dmytruk_hybrid_2024,kass2024manybodyphotonblockadequantum} as well as ab-initio methods~\cite{ruggenthaler2014quantum,flick2018cavity}. Numerical methods based on matrix product states have been extended to include coupling to photons~\cite{bacciconi2023first,bacciconi2025theory,shaffer_entanglement_2023}. Another direction involves deriving effective electron-only Hamiltonian after integrating out the photonic mode, either exactly when possible or perturbatively, in the case electron and photon are highly off-resonant. This strategy has been pursued for example in Refs.~\cite{ciuti2021cavity,arwas2023quantum,andolina_amperean_2024} via adiabatic elimination of the cavity or in Refs.~\cite{sentef_quantum_2020,li_effective_2022,perez-gonzalez_light-matter_2025,becerra2025fermionparityswitchesimprinted} within the so called quantum-Floquet framework 
which uses approximate block-diagonalization approaches to obtain the effective electronic Hamiltonian. The latter amounts to write down the electronic Hamiltonian in the Fock-basis of the cavity modes and interpret the resulting matrix elements in analogy with Floquet theory of driven quantum systems.

%The theoretical description of electrons strongly coupled to light has been done in various ways.
%In Ref. \cite{mazza_superradiant_2019}, an extensive action is derived and a saddle-point analysis is performed. However, in the absence of a phase transition for the photons in the cavity, this approach fails to capture the effect of the cavity on matter.
%Using a mean-field decoupling of light and matter like in Ref. \cite{dmytruk_controlling_2022} the effect of the cavity is only to renormalize the parameters of the electrons' Hamiltonian but do not generate interactions like we expect. The gaussian fluctuations can be added around the mean-field to obtain an effective description of the cavity in the form of the polaritonic spectrum like in Ref. \cite{dmytruk_controlling_2022,dmytruk_hybrid_2024}. However, this approach relies on integrating out the electrons and thus makes their description impossible. 

In this work, we will derive an effective Hamiltonian for the electrons coupled to an off-resonant cavity using Floquet theory and the high-frequency expansion~\cite{goldman_periodically_2014,eckardt_high-frequency_2015,Bukov04032015}. For equilibrium problems, i.e. a cavity-electron system in the ground-state, this can be done naturally by going into a rotating frame at the frequency corresponding to the high-energy scale and perform the high-frequency expansion on the resulting Floquet problem~\cite{bukov_schrieffer-wolff_2016}. We show that the resulting electron-only Hamiltonian contains cavity-mediated interactions, a genuine feature of treating the light field quantum mechanically. While our approach is perturbative in the inverse of the frequency of the cavity mode, it treats light-matter coupling non-perturbatively within the Peierls phase, including multiphoton processes. In addition, this approach will naturally allow us to treat driven electron-photon problems, such as in presence of cavity or electron driving.

We apply our framework to study a Su-Schrieffer-Heeger (SSH) model coupled to a single cavity mode, a model which has received attention recently in the context of cavity-induced topology~\cite{dmytruk_controlling_2022,perez2022topology,shaffer_entanglement_2023,nguyen2024electron,perez-gonzalez_light-matter_2025,zhao2025observation}. By using our framework we show that corrections to high-frequency limit result in modification of the topological phase diagram that we ascribe to the presence of light-matter entanglement. Our results further highlights the many-body nature of the strong light-matter coupling regime, where first order corrections to the HFE at fixed frequency can become dominant and push the system away from the single particle regime.
%The effective Hamiltonian obtained displays cavity-mediated interactions. In the literature, those effective electron-electron interactions mediated by the cavity have been captured via adiabatic elimination of the cavity~\cite{andolina_amperean_2024} and in the so-called quantum Floquet picture~\cite{li_effective_2022}. The interaction derived in Ref. \cite{andolina_amperean_2024} will be shown to coincide with ours, while the formalism detailed in Ref. \cite{li_effective_2022} will be shown to reduce to ours for a single-mode cavity. However, our approach has the advantage of being easily computed as it relies on the Fourier decomposition rather than on the projection on photon sectors. Furthermore, the example detailed in Sect. \ref*{sec:SplitRing} illustrates that the physics generated by a single-mode cavity differs vastly from that of multimode cavity such as those studied in Ref. \cite{li_effective_2022}.

%\textcolor{green}{I have to mention that I don't truncate the Peierls phase and I consider all $m$ photons processes. And frame the article as a deep dive in the connection between Floquet and Photons.}

%\textcolor{green}{I have to mention that my results highlight that at high coupling and at a fixed frequancy the first order can have an effect and even be dominant. Thus the zeroth order truncation of sentef and Alvaro may be very wrong : the problem is really may-body at high coupling, with our formalism leading to an easy treatement. }
The manuscript is structured as follows. In Sec.~\ref{sec:HFEgeneralprogram} we introduce our cavity-electron set-up and formulate it in term of an effective Floquet problem. In Sec.~\ref{sec:HFE} we derive an effective Hamiltonian for the electrons strongly coupled to a high-frequency single-mode cavity. Then, in Sec.~\ref{sec:VanVleck} we connect our formalism to results in the literature, such as Van-Vleck perturbation theory and quantum Floquet picture. In Sec.~\ref{sec:DrivenHFE} we extend our Floquet approach to the case of a driven cavity or electronic system. Finally, in Sec.~\ref{sec:SSH} we present an application to the SSH model coupled to a cavity. Sec.~\ref{sec:conclusions} is devoted to conclusions, while several Appendixes complete the work with further theoretical details.

%Then, in Sec.~ \ref{sec:gaugeInvariant} the gauge invariance of the previously derived scheme is investigated. Finally, Sect. \ref{sec:SplitRing} is dedicated to the application of the formalism to the graphene embedded in a split-ring resonators. Details of the calculations can be found in appendix. 

\section{Model and Effective Floquet Picture}\label{sec:HFEgeneralprogram}

We will consider a  fermionic lattice system, described by the set of fermionic operators $\mathbf{c}^{\dagger}$, $\mathbf{c}$, coupled to a \emph{far off-resonant} single-mode cavity described by a bosonic mode $a^{\dagger}$, $a$. The off-resonant regime corresponds to the cavity frequency $\omega_c$ being the dominant energy scale of the problem. Thus, generally the Hamiltonian for the problem will be of the form :
\begin{equation}
    \hat{H} = \hat{H}_{el}[a,a^{\dagger}, \mathbf{c}, \mathbf{c}^{\dagger}] + \underbrace{\hbar \omega_c a^{\dagger}a}_{\text{dominant term}} .
\end{equation}
In what follows, we will study first the case of a dark cavity, namely a cavity with no illumination, where the cavity electromagnetic field is in its vacuum state. Even though there are no photons in the cavity, this case is non trivial as the vacuum fluctuations of the quantum electromagnetic field may have an effect on the electrons. 

\subsubsection{The tight-binding model coupled through Peierls phase}

The electrons on a lattice are described by a tight-binding model characterized by the hopping integrals from site $i$ to site $j$ denoted by $t_{i,j}$. In all generality, the index $i$ regroups the lattice site $\vec{R}_i$, the spin $\sigma$ and if needed sub-lattice index $\kappa$. Formally, the index $i$ is a triplet : $i = (\vec{R}_i,\sigma,\kappa)$. Sums will run over all the sub-indices and are restricted by the form of the hopping integrals. For example, if the electrons conserve their spin when hopping, the hopping integrals are of the form $t_{i,j} = t_{\vec{R}_i, \kappa, \vec{R}_j, \kappa^{\prime}}\delta_{\sigma, \sigma^{\prime}}$. The uncoupled electrons Hamiltonian is given by : 
\begin{equation} \label{eq:uncoupledTBmodel}
    \hat{H}_{el} =  \sum_{i,j} t_{i,j}c_i^{\dagger}c_j + \hat{H}_{int} ,
\end{equation}
where $c_i^{\dagger}$ and $c_i$ are the fermionic creation and destruction operators of the electrons associated to a certain basis and $\hat{H}_{int}$ contains the electron-electron interactions and does not couple light and matter. 

The cavity hosts a single mode of frequency $\omega_c$ whose vector potential is written $\hat{\vec{A}}(\vec{r}) = \vec{A}_0(\vec{r}) a^{\dagger} + \vec{A}_0^{\ast}(\vec{r}) a$ with $a^\dagger$ and $a$ the bosonic creation and annihilation operators of the cavity mode.
Matter is coupled to light via Peierls substitution so that the Hamiltonian of matter coupled to the cavity mode is :
\begin{equation}\label{eq:InitialTightBindingHamiltonian}
    \hat{H} = \omega_c a^{\dagger} a + \sum_{i,j} t_{i,j} e^{-\mathrm{i}(g_{i,j}a^{\dagger} + g_{i,j}^{\ast} a  ) } c_i^{\dagger}c_j + \hat{H}_{int} ,
\end{equation}
where the Peierls phases $g_{i,j}$ are defined through :
\begin{equation}\label{eq:defPeierlsPhases}
    g_{i,j} \equiv \frac{e}{\hbar} \int_{\vec{R}_i}^{\vec{R}_j} {\vec{A}_0}({\vec{r}})\cdot \vec{dr}  ,
\end{equation}
and only depend on the lattice sites $\vec{R}_i$ and $\vec{R}_j$.
% \sout{\textcolor{blue}{In the following we disregard for simplicity the spin of the electrons and treat them as spinless fermions.}}
% \textcolor{red}{MS: Peierls phase for multiorbital systems is a delicate issue. Plus the lattice site has a specific meaning in the Peierls phase and does not include the spin. So either you write explicitly the spin in the electron operator in Eq. 6, or you need a sentence saying that you drop the spin for simplicity.}
%\textcolor{red}{MS: as also mentioned later, since we never discuss $H_{int}$ we might also remove it from the Hamiltonian.}\textcolor{green}{I add forgotten to make the remark but i have now added it : the cavity-meditaed interaction are added on top of the initial interactions of the model.}

A remark can be made that the relevant energy scale of the electrons is set by $t_{i,j}\sim t$ and that the coupling to light is always of order one. As such, the High Frequency expansion will be perturbative in the small parameter $\delta =( {t}/{\omega_c})$ rather than in the light-matter coupling strength. %Thus, starting from the Peierls phase gauge {\color{red} ?} allows to derive an effective Hamiltonian which is perturbative in $\frac{1}{\omega_c}$ but not in the light-matter coupling strength. This may not be the case in other settings or other gauges and will be explored more in Sect. \ref{sec:gaugeInvarian}.
% Make a remark that it is non perturbative in g but just in this gauge

Hence, we want an effective Hamiltonian describing the electrons which incorporates the effects of the vacuum fluctuations of the cavity mode, in the regime where this is off-resonant with respect to the electronic transitions.

To proceed we follow a strategy introduced in the context of periodically driven quantum systems. For large driving frequency energy absorption is suppressed and the dynamics is controlled by an effective \emph{Floquet} Hamiltonian which can be constructed perturbatively in an expansion in the inverse of the drive frequency ~\cite{goldman_periodically_2014,eckardt_high-frequency_2015}. This construction turns out to be fully equivalent to the equilibrium problem of deriving an effective Hamiltonian by Schrieffer-Wolff transformation, eliminating high-energy degrees of freedom and renormalizing the low-energy ones. The analogy can be made precise by going in a rotating frame at the frequency corresponding to the high-energy scale and perform the high-frequency expansion on the resulting Floquet problem~\cite{bukov_schrieffer-wolff_2016}. Here we follow this approach in the context of our cavity-electron system in Eq.~(\ref{eq:InitialTightBindingHamiltonian}). To describe the low energy physics, the dominant term is eliminated by going into a rotating frame via the unitary transformation :
\begin{equation}\label{eq:defRotatingFrame}
    \hat{W}(t) = e^{-i \omega_c a^{\dagger}a t},
\end{equation}
so that the Hamiltonian in the rotating frame is $\hat{H}(t) =\hat{W}^{\dagger} \hat{H}\hat{W} + i \left(\partial_t \hat{W}^{\dagger}\right)\hat{W}$. Then using the boson commutation relation, one obtains $\hat{W}^{\dagger}a\hat{W} = ae^{-i\omega_c t}$, so that the Hamiltonian $\hat{H}$ reads :
\begin{equation}\label{eqn:H_timedep}
    \hat{H}(t) = \hat{H}_{el}[ae^{-i\omega_c t},a^{\dagger}e^{i\omega_c t},\mathbf{c}, \mathbf{c}^{\dagger}] .
\end{equation}
It is clear that $\hat{H}(t)$ is now $T=({2\pi}/{\omega_c})$-periodic thus \emph{Floquet theory} can be used in the high-frequency regime $\omega_c\rightarrow\infty$ to obtain long-time, i.e. low energy, dynamics of the system through a time-independent so-called Floquet Hamiltonian $\hat{H}_F$~\cite{goldman_periodically_2014,Bukov04032015,dalibard_periodically_2015} as we are going to discuss below.
%\textcolor{red}{MS: we should say something about what to do with HF then..(especially for driven cavity)}

\section{High-Frequency expansion}\label{sec:HFE}

In the high-frequency regime we can obtain the Floquet Hamiltonian of the system within perturbative expansion in $({1}/{\omega_c})$ called the High Frequency Expansion (HFE). The details and results of the derivation of such an expansion can be found in \cite{eckardt_high-frequency_2015}.  The expansion is described in terms of the Fourier coefficients of the time-periodic Hamiltonian in Eq.~(\ref{eqn:H_timedep}) :
\begin{equation}\label{eq:defFourierCoeffs}
    \hat{H}_m \equiv \frac{1}{T} \int_0^T \hat{H}(t) e^{-i m\omega_c t}\ dt ,
\end{equation}
so that to first order, the Floquet Hamiltonian reads~\cite{eckardt_high-frequency_2015}
\begin{equation} \label{eq:FloquetHFEgeneral}
\hat{H}_F = \hat{H}_{m=0} + \sum_{m>0}\frac{
[\hat{H}_{m},\hat{H}_{-m}]}
{m\omega_c} + \mathcal{O}(\frac{1}{\omega_c^2}) .
\end{equation}
We now compute the effective Floquet Hamiltonian of our electron-photon system via the HFE. First, going into the rotating frame gives the time-dependent Hamiltonian 
\begin{equation}\label{eq:rotatingTightBindingHamiltonian}
    \hat{H} = \sum_{i,j} t_{i,j} e^{-i(g_{i,j}e^{i\omega_c t }a^{\dagger}  + g_{i,j}^{\ast}e^{-i\omega_c t } a  ) } c_i^{\dagger}c_j + \hat{H}_{int}.
\end{equation}
for which the Fourier coefficients takes the form 
\begin{equation} \label{eq:FourierCoeffTB_Gamma}
    \hat{H}_m = \sum_{i,j}  t_{i,j} \hat{\gamma}_{m}(g_{i,j}) \otimes c_i^{\dagger}c_j + \delta_{0,m} \hat{H}_{int}\,.
\end{equation}
Here the operators $\hat{\gamma}_{m}(g)$ act on the photon Hilbert space and are defined as
\begin{equation}\label{eq:DefGammaTB}
    \hat{\gamma}_{m}(g) \equiv \frac{1}{2\pi}\int_0^{2\pi} e^{ - i (g e^{i \omega_c t} a^{\dagger} + g^{\ast}e^{-i \omega_c t} a)  }e^{-i m t}\ dt\,.
\end{equation}
For $m\geq0$ we can obtain an expression for $\hat{\gamma}_{m}(g)$ of the form
\begin{equation} \label{eq:GammaExpression}
    \hat{\gamma}_{m}(g) = e^{- |g|^2/2 } \sum_{n\geq 0} \frac{(-i)^{2n + m }  {g}^{n+m} {g^{\ast} }^{n} }{(n+m)! n! }{a^{\dagger} }^{n+m}a^n
\end{equation}
from which we see that $\hat{\gamma}_{-m}(g) = \left(\hat{\gamma}_{m}(-g) \right)^{\dagger}$.
% \begin{widetext}
% \begin{equation}\label{eq:GammaExpression}
%     \hat{\gamma}_{m}(g) = \left\{\begin{array}{c}
%     e^{- |g|^2/2 } \sum_{n\geq 0} \frac{(-i)^{2n + m }  {g}^{n+m} {g^{\ast} }^{n} }{(n+m)! n! }{a^{\dagger} }^{n+m}a^n \text{ if }m\geq 0\\
%     e^{- |g|^2/2  } \sum_{n\geq 0} \frac{(-i)^{2n + |m| }  {g}^{n} {g^{\ast} }^{n+|m|} }{(n+|m|)! n! }{a^{\dagger} }^na^{n+|m|} \text{ if }m\leq 0\\    
%     \end{array}\right. .
% \end{equation}
% \end{widetext}
We note that the Fourier coefficient $\hat{H}_{m}$ regroups all processes which create algebraically $m$ photons.
% The detailed calculations can be found in appendix \textcolor{red}{Shall we add details in the appendix?}
From Ref.~\cite{eckardt_high-frequency_2015}, the zeroth order (in $({1}/{\omega_c})\ $) Floquet Hamiltonian is given by :
\begin{equation*}
    H_F^{(0)} = \hat{H}_{m=0},
\end{equation*}
so that using Eq.~(\ref{eq:GammaExpression}) yields the same result as that of Ref.~\cite{sentef_quantum_2020}: 
\begin{equation}\label{eq:zeroOrderFloquet}
    H_F^{(0)} = \sum_{i,j} t_{i,j} \left( e^{- |g_{i,j}|^2/2  } \sum_{n\geq0}   \frac{(-1)^n  |g_{i,j}|^{2n}}{(n!)^2}\ {a^{\dagger} }^{n}a^n  \right)\otimes c_i^{\dagger}c_j .
\end{equation}
This photon-conserving Hamiltonian can be interpreted as the dressing of the hopping of electrons by processes with $n$ photons.
% This result can be interpreted as the dressing of the hopping of electrons between sites by processes where the hopping between two sites is mediated by $n$ photons. Furthermore, the Floquet Hamiltonian is photon-conserving.
The first order (in $({1}/{\omega_c})$) Floquet Hamiltonian is given by :
\begin{equation}\label{eq:formulaFloquetHFEatFirstOrder}
    H_F^{(1)} = \sum_{m>0 }\frac{1}{m\omega_c}[\hat{H}_{m},\hat{H}_{-m}].
\end{equation}
% Which, using Eq. \ref{eq:FourierCoeffTB_Gamma}, can be written as :
% \begin{align}
%     H_F^{(1)} &= \sum_{m>0 }\frac{1}{m\omega_c} \sum_{i,j,k,l} t_{i,j}t_{k,l} [ \hat{\gamma}_{m}\left(g_{i,j}\right)\otimes c_i^{\dagger}c_j,\hat{\gamma}_{-m}\left(g_{k,l}\right)\otimes c_k^{\dagger}c_l].
% \end{align}
It has to be noted that the commutator of tensor product operators is not the tensor product of the commutators. Instead, the structure of the tensor product Hilbert space imposes 
\begin{equation}\label{eq:commutatorIdentityTensorProduct}
    [A\otimes B ,C\otimes D ] = [A,C]\otimes BD + CA\otimes[B,D].
\end{equation}
By making use of the following identity derived from the fermionic commutation relations 
\begin{equation}\label{eq:CommutaionRelatioHoppingFermions}
    [c_i^{\dagger}c_j, c_k^{\dagger}c_l] = \delta_{k,j} c_i^{\dagger} c_l- \delta_{i,l}c_k^{\dagger}c_j,
\end{equation}
the first-order Floquet Hamiltonian can be divided into two terms: 
\begin{equation}
    H_F^{(1)} = H_{F,hop}^{(1)}+H_{F,int}^{(1)},
\end{equation}
respectively describing cavity-induced hopping and electron-electron interactions, which read after proper normal-ordering 
\begin{widetext}
\begin{align}
    H_{F,hop}^{(1)} &=   \sum_{i,j}\parent{\sum_{m\neq 0}\frac{1}{m\omega_c} \sum_l t_{i,l}t_{l,j}\ \hat{\gamma}_{m}(g_{i,l})  \hat{\gamma}_{-m}(g_{l,j}) }\otimes c_i^{\dagger} c_j \label{subeq:FirstOrderHopping} \\
    H_{F,int}^{(1)} &= \sum_{i,j,k,l} t_{i,j}t_{k,l} \parent{\sum_{m>0 }\frac{1}{m\omega_c} \commut{\hat{\gamma}_{m}\left(g_{i,j}\right)}{\hat{\gamma}_{-m}(g_{k,l}) }} \otimes c_i^{\dagger} c_k^{\dagger}c_l c_j \label{subeq:FirstOrderInteractions} .
\end{align}
\end{widetext}

The cavity-mediated hopping and interactions are composed of all $m$-photons second-order processes. However, both the interactions and hopping part conserve the number of photon.

Furthermore, the cavity-mediated electron-electron interactions appear as a direct consequence of the quantum nature of the electromagnetic field. Indeed, they are made up of commutators of (functions of) the electromagnetic vector potential $\hat{A}$ which trivially vanish if the vector potential is not an operator i.e. if the field is classical. Finally, we note that at first order the cavity-mediated interactions are added up on top on the initial interactions in the model.

It has to be pointed out that cavity-mediated hopping term can be absorbed into the interaction if the latter are not normal ordered. Indeed, starting from an equivalent formula for the Floquet Hamiltonian, given by
\begin{equation}\label{eq:FirstOrderFloquetHamiltonianWithoutCommutator}
    \hat{H}_{F}^{(1)} = \sum_{m\neq 0 }\frac{\hat{H}_{m}\hat{H}_{-m}}{m\omega_c},
\end{equation}
it is clear that only an interaction term appears, at the cost of being not normal ordered. While this form is simpler for calculations, and we will make use of it extensively in the remaining of the draft, it lacks the physical significance of the first one, as highlighted by the classical limit that we will discuss below.

\subsection{Examples}

The effective Floquet Hamiltonian derived above to leading order in the HFE still contains both electrons and cavity degrees of freedom. By projecting this Hamiltonian on a given cavity state we can obtain the desired effective electronic Hamiltonian describing cavity-mediated processes. Here we provide explicit examples for a cavity in the vacuum or in a coherent state.

\subsubsection{Vacuum cavity}
% \textcolor{green}{Do I talk about and plot the n photon case? Or just the $n=0$.} 

Since the Floquet Hamiltonian is photon conserving, the projection on a fixed photon number sector is exact. On the $n$ photon sector, the Floquet Hamiltonian of Eq.~\ref{eq:FirstOrderFloquetHamiltonianWithoutCommutator} to order one in the HFE can be put in the form  
\begin{equation} \label{eq:NphotonSector_FirstOrderHamiltonian}
\hat{H}_{F,n}^{[1]} = \sum_{i,j}t^{\rm eff}_{i,j}
c_i^{\dagger} c_j + \sum_{i,j,k,l}V^{\rm eff}_{ijkl}c_i^{\dagger}  c_j c_k^{\dagger}c_l,
\end{equation}
where the effective, cavity-mediated, hopping and interactions read respectively $t^{\rm eff}_{i,j}=t_{i,j}\mathcal{I}_{n}(g_{i,j})$ and $V^{\rm eff}_{ijkl}=({t_{i,j}t_{k,l}}/{\omega_c})\mathcal{K}_{n}(g_{i,j},g_{k,l})$. Here the functions $\mathcal{I}_n(g)$ and $\mathcal{K}_n(g,g')$ are defined as
\begin{align}
\mathcal{I}_n(g)&= \left<n\right|\hat{\gamma}_0(g)\left|n\right>\\
\mathcal{K}_{n}(g, g^{\prime})&=\sum_{m\neq0}\frac{\left<n\right|\hat{\gamma}_{m}(g)\hat{\gamma}_{-m}(g^{\prime})\left|n\right> }{m}    
\end{align}
Expressions for these functions can be obtained from Eq.~(\ref{eq:GammaExpression}) and, while cumbersome, allow the numerical evaluation of the coefficients of the model.

\begin{figure}[t]
        \centering  
        \includegraphics[width=\linewidth]{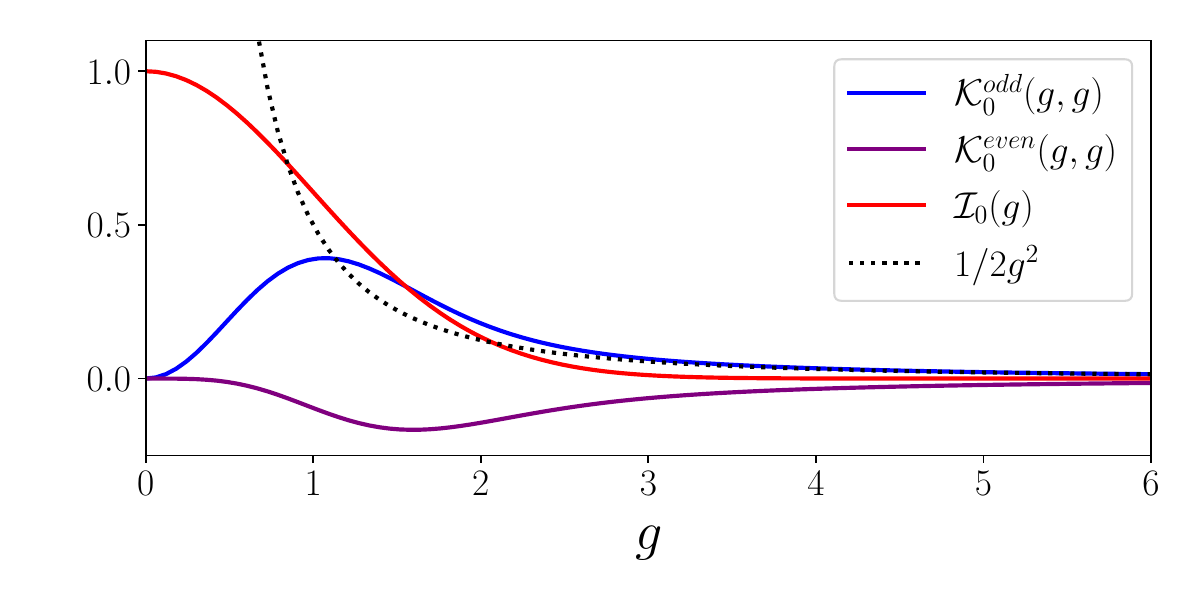}
        \caption{\justifying Plot of the coefficients appearing in the vacuum cavity effective Hamiltonian. $ \mathcal{I}_{0}(g )$ controls the zero order and accounts for the exponential renormalization of the hopping term. $\mathcal{K}_{0}^{even,odd}(g, g)$ control respectively the density-density and current-current interactions, see Eq.~(\ref{eq:FloquetHamiltonianVacuumCurrentHoppingDecomposition}).
        \label{fig:CoefficientsOddEven} }
\end{figure} 
Because the cavity mode energy $\omega_c$ is large when compared to every other energy scale, and because we are looking at an equilibrium problem (no driving), there is a strong case for the cavity to be in the vacuum state $\ket{0}$ in the ground state of the system. In the case of a vacuum cavity, the coefficients read 
$$ \mathcal{I}_{0}(g ) = e^{-\left|g \right|^2/2},$$
$$ \mathcal{K}_{0}(g, g^{\prime})   = e^{-\frac{ \left|g\right|^2+ \left|g^{\prime}\right|^2   }{2}}\int_{0}^{ g{g^{\prime}}^{\ast}}\frac{1-e^{-s} }{s}\ ds , $$
Upon the definition of the current and hopping operators respectively through 
\begin{align*}
\hat{K}_{i,j} &\equiv \frac{1}{2} t_{i,j}c_i^{\dagger}c_j + {\rm h.c}\ ,\\
\hat{J}_{i,j} &\equiv \frac{\ii}{2}t_{i,j}c_i^{\dagger}c_j + {\rm h.c}\ , 
\end{align*}
the interactions can be put in the form 
\begin{align}
\hat{H}_{F,0}^{(1)} &= \sum_{i,j,k,l}  \mathcal{K}_{0}^{ even}(g_{i,j},g_{k,l}) \hat{K}_{i,j}\hat{K}_{k,l}\notag \\
& -\sum_{i,j,k,l}\mathcal{K}_{0}^{odd}(g_{i,j},g_{k,l}) \hat{J}_{i,j}\hat{J}_{k,l},\label{eq:FloquetHamiltonianVacuumCurrentHoppingDecomposition}
\end{align}
where the coefficients have been decomposed into their odd and even parts $\mathcal{K}_{0}^{ even/odd}(g,g^{\prime}) \equiv \left[\mathcal{K}_{0}(g,g^{\prime}) \pm \mathcal{K}_{0}(g,-g^{\prime})   \right]/2$. The diagonal components of these functions are displayed in Fig.~\ref{fig:CoefficientsOddEven}. These show that the cavity mediates attractive hopping-hopping interactions and amperean current-current interactions. Furthermore, at very high coupling strengths, the interactions dominate with respect to the cavity-renormalized hopping as the former decays as ${g^{-2}}$ while the latter decays as $e^{-\left|g\right|^2/2}$. This relatively slow decay is a direct consequence of the absence of truncation of the Peierls phase.

On the other hand, at low coupling $\mathcal{K}_{0}^{ even}$ scales as $g^4$ while $\mathcal{K}_{0}^{odd}$ grows as $g^2$ since the former originates from the diamagnetic and higher even order of the Peierls phase. As such, in all theories considering only a linear light-matter coupling, this term drops out and only current-current interactions are mediated by the cavity, possibly disturbing the hierarchy of the terms detailed in Fig.~\ref{fig:tableCoeff}. 

Finally, it is worth noting that once the projection has been performed, the system is described purely by an electronic Hamiltonian and can be treated using the usual techniques, such as Hartree-Fock.

% \textcolor{green}{To do : plot of the $\mathcal{K}_{0}(g,\pm g)$ and/or do the current-hopping decomposition.}

% The following remarks can be made on the integral appearing above : 
% \begin{subequations} \label{eq:propertiesOfIntegral}
% \begin{align}
% \forall x\in\R, \ \text{sgn}(&\int_{0}^x \frac{1 - e^{-s} }{s}\ ds) = \text{sgn}(x) \\
% &\int_{0}^x \frac{1 - e^{-s} }{s}\ ds \underset{x\rightarrow0}{\sim } x
% \end{align}
% \end{subequations}

\subsubsection{Vacuum cavity in the thermodynamic limit}

It is interesting to discuss the effect of cavity mediated processes in the thermodynamic limit,  where the system size $V\longrightarrow \infty$. In this regime the light-matter coupling typically scales as $g_{i,j}\sim 1/\sqrt{V}$. In this limit, the behavior of the cavity-mediated interaction detailed above show that the leading term is given by the current-current interaction of the form :
\begin{equation} \label{eq:FirstOrderHamiltonian_ThermoLimit}
\hat{H}_F^{[1]} = \hat{H}_{el} - \frac{\hat{J}\hat{J}^{\dagger}}{\omega_c {V}},
\end{equation}
where $\hat{H}_{el}$ is the bare electrons Hamiltonian defined in Eq.~(\ref{eq:InitialTightBindingHamiltonian}) and $\hat{J}$ is a generalized current operator defined through : 
\begin{align*} 
    \hat{J} \equiv \frac{\ii}{2} \sum_{i,j}\left( t_{i,j}  g_{i,j}^{\ast} \sqrt{V} c_i^{\dagger}c_j - t_{i,j}^{\ast} g_{i,j} \sqrt{V} c_j^{\dagger}c_i\right).
\end{align*}
This operator is extensive, hence the Floquet Hamiltonian scales extensively. On can easily checks that for a plane wave cavity mode i.e. $A_{0}(\vec{r}) \propto e^{\ii \vec{q}\cdot\vec{r}} $, the generalized current $\hat{J}$ is proportional to the paramagnetic current at wave-vector $\vec{q}$. Thus, at the leading order in the thermodynamic limit for the electrons, the off-resonant cavity only mediates amperean current-current interaction with infinite range.

For sufficiently high light-matter coupling, the Hamiltonian of Eq.~\ref{eq:FirstOrderHamiltonian_ThermoLimit} predicts the forbidden superradiant phase transition \cite{andolina_cavity_2019} within a mean-field treatement like that of Ref.~\cite{lenk_collective_2022} which becomes exact in the thermodynamic limit. Indeed, taking the thermodynamic limit at this step is equivalent to truncating the Peierls phase coupling at first order. This operation is known to break the gauge invariance which normally forbids the phase transition. As such, care as to be taken to conserve the gauge invariance in approximation schemes \cite{dmytruk_gauge_2021,li_electromagnetic_2020,de_bernardis_breakdown_2018,di_stefano_resolution_2019}. 
%The issue of the preservation of the gauge invariance within the HFE is discussed in App.~\ref{sec:GaugeEquivalence}.

\subsubsection{Classical limit}\label{subsubsec:classicalLimit}

We now consider the classical limit of a cavity in a coherent state with many photons, as discussed in Ref.~\cite{sentef_quantum_2020}. We expect that in this regime the cavity-mediated interactions discussed in the previous section would vanish, since it is known that for  Floquet driven non-interacting electrons a classical drive can only mediate long-range hopping~\cite{oka_floquet_2019}. The cavity is supposed to be in a coherent state with no light-matter entanglement, so that $\ket{\Phi} = \ket{\alpha}\otimes\ket{\psi_{el}}$. Then, the confinement of the electromagnetic field is sent to zero $V_{mode}\longmapsto \infty$ while the energy density in the cavity (or equivalently the amplitude of the electric field) is kept constant. Furthermore, the one-photon vector potential in the cavity scales as $A_0 \propto \frac{1}{\sqrt{V_{mode}}}$ while the energy stored in the electromagnetic field scales as $\mathcal{E}_{elm}\propto V_{mode}$.
The first consideration imposes that $g_{i,j} \propto \frac{1}{\sqrt{V_{mode}}}$ while the second gives $|\alpha|\propto \sqrt{V_{mode}}$. As such, the classical limit in the present context is defined as :
\begin{equation*}
    \left\{ \begin{array}{c}\alpha \longrightarrow\infty\\g_{i,j}\alpha = c_{i,j} \text{ fixed}\end{array}\right.
\end{equation*}

In this case, the effective Hamiltonian for the electrons at zeroth order reads 
\begin{equation*}
    \hat{H}_{eff,el}^{(0)} = \sum_{i,j}  t_{i,j} { e^{- |g_{i,j}|^2 /2 } } J_0(2|\alpha g_{i,j}|)  c_i^{\dagger}c_j.
\end{equation*}
where $\mathcal{J}_0$ stands for the zeroth order Bessel function of the first kind. 
In the classical limit, this expression converges to the well-known case of classically-driven electrons on a lattice with an electric field amplitude of $E_0 = \omega_c A_0 \alpha$ \cite[eq. 131]{eckardt_high-frequency_2015}, thus, recovering the result of Ref.~\cite{sentef_quantum_2020}.

The first order Hamiltonian with the cavity in a coherent state can be written as
\begin{align*} 
    \hat{H}_{F,hop}^{(1)}(\alpha)&=  \sum_{i,j}\mathcal{T}_{i,j}(\alpha)  c_i^{\dagger} c_j ,\\
    \hat{H}_{F,int}^{(1)}(\alpha) &= \sum_{i,j,k,l}\frac{t_{i,j}t_{k,l}}{\omega_c}\mathcal{U}_{i,j,k,l}(\alpha) c_i^{\dagger}c_k^{\dagger}c_l c_j ,
\end{align*}
where the coefficients of the cavity-mediated hopping and interaction are given by the following formulas in terms of higher-order Bessel functions
\begin{widetext}
\begin{align}
    \mathcal{T}_{i,j}(\alpha) &= \sum_{m\neq 0}\sum_{l} \sum_{n\geq 0}(-1)^m\frac{t_{i,l} t_{l,j } }{m \omega_{cav} }e^{ -\frac{|g_{i,l}|^2 + |g_{l,j}|^2 }{2} } e^{i m \  \text{arg}\parent{\frac{g_{i,l}}{g_{l,j}}}  } {\frac{|g_{i,l}g_{l,j}|^n}{n!} }  \mathcal{J}_{n+m}( 2 |g_{i,l}\alpha|  )\mathcal{J}_{n+m}( 2 |g_{l,j}\alpha|  ),\label{eq:CoherentStateHopping} \\
    \mathcal{U}_{i,j,k,l}(\alpha) &= e^{-\frac{ \left|g_{i,j}\right|^2+ \left|g_{k,l}\right|^2   }{2}} \sum_{m>0}(-1)^m\frac{e^{\ii m \ \text{arg}\parent{\frac{g_{i,j}}{g_{k,l}}}  } }{m}\sum_{n\geq 0} \frac{{|g_{i,j}g_{k,l}|}^n}{n!} %\notag  \\ & \times 
    \left(  \mathcal{J}_{m+n}(2|g_{i,j}\alpha|) \mathcal{J}_{n+m}(2|g_{k,l}\alpha|) \right. \notag  \\ & \left.-\mathcal{J}_{n-m}(2|g_{i,j}\alpha|) \mathcal{J}_{n-m}(2|g_{k,l}\alpha|)  \right)   .\label{eq:CoherentStateInteraction}  
\end{align}
\end{widetext}
% Finally, it is useful to discuss the classical limit. As stated earlier, the cavity-mediated interactions term only appear because of the quantum nature of the field in the cavity as they are proportional to a commutator. One thus expects that within a "classical limit", the interactions vanish and the results obtained for a classically driven system are recovered.

% \begin{widetext}
% \begin{align*}
%     \mathcal{U}_{i,j,k,l}(\alpha)% &= e^{-\frac{ \left|c_{i,j}\right|^2+ \left|c_{k,l}\right|^2   }{2|\alpha|^2}} \sum_{m>0}(-1)^m\frac{e^{\ii m \  \text{arg}\parent{\frac{c_{i,j}}{c_{k,l}}}  } }{m}\sum_{n\in\N} \frac{{|c_{i,j}c_{k,l}|}^n}{n!|\alpha|^{2n}}\parent{  \mathcal{J}_{m+n}(2|c_{i,j}|) \mathcal{J}_{n+m}(2|c_{k,l}|) -\mathcal{J}_{n-m}(2|c_{i,j} |) \mathcal{J}_{n-m}(2|c_{k,l}|)  } \\
%     &= e^{-\frac{ \left|c_{i,j}\right|^2+ \left|c_{k,l}\right|^2   }{2|\alpha|^2}} \sum_{m>0}(-1)^m\frac{e^{\ii m \  \text{arg}\parent{\frac{c_{i,j}}{c_{k,l}}}  } }{m}\parent{  \mathcal{J}_{m}(2|c_{i,j}|) \mathcal{J}_{m}(2|c_{k,l}|) -\mathcal{J}_{-m}(2|c_{i,j} |) \mathcal{J}_{-m}(2|c_{k,l}|)  }  +\frac{1}{|\alpha|^2} \sum_{n\geq1}  \dots
% \end{align*}
% \end{widetext}
% so that a $1/|\alpha|^2$ is factorized from the $n\geq1$ sum, thus prooving its vanishing in the thermodynamic limit.  leaving us with :
In the classical limit, both Eq.~(\ref{eq:CoherentStateHopping}) and Eq. ~(\ref{eq:CoherentStateInteraction}) can be split in the $n=0$ term, which does not contain powers of the light-matter coupling $g_{i,j}$, and the remaining terms for $n\neq0$ where a dependence $\frac{1}{|\alpha|^2}$ can be factor out. However, due to the symmetry of Bessel functions and the structure of the sum inherited from the commutator of Eq.~\ref{subeq:FirstOrderInteractions}, the $n=0$ term in the expression for the interaction $U_{i,j,k,l}(\alpha)$ vanishes, while the same term for the hopping remains finite. As such, in the classical limit the cavity-mediated interactions and hopping reduce to 
\begin{widetext}
    \begin{align*}
        \mathcal{T}_{i,j}(\alpha) &= \sum_{m\neq 0}\sum_{l} (-1)^m \frac{t_{i,l} t_{l,j } }{m \omega_{cav}}e^{\ii m \  \text{arg}\parent{\frac{c_{i,l}}{c_{l,j}}}  }   \mathcal{J}_{m}( 2 |c_{i,l}|  )\mathcal{J}_{m}( 2 |c_{l,j}|  ) +\mathcal{O}(\frac{1}{|\alpha|^2}) \text{\kern+2em and \kern+2em}\mathcal{U}_{i,j,k,l}(\alpha) = \mathcal{O}(\frac{1}{|\alpha|^2}).
\end{align*}
\end{widetext}
We note that the non-vanishing term in the cavity-mediated hopping coincides with the result of the Floquet with classical light such that the classical Peierls phase associated to the driving field are given by $c_{i,j}$. Furthermore, when $g_{i,j}\in\R$ the term vanishes. This is the case too for the classical drive, and is the reason why circularly polarized light is needed to obtain a Haldane model when shined on graphene in Ref. \cite{eckardt_high-frequency_2015,mciver_light-induced_2020,oka_floquet_2019}.

\section{Link to Van Vleck perturbation theory and Quantum Floquet Theory}\label{sec:VanVleck}

In the previous derivation of the effective Hamiltonian, the use of Floquet theory is just a mathematical procedure with the time-dependence of $\hat{H}$ having no physical significance and appearing purely because of the unitary transformation performed to remove the dominant term. However, this artificial time-dependence blurs the concept of thermal equilibrium, namely because of the micromotion operator which could lead to a steady-state where quantities still oscillate at the frequency $\omega_c$.  

Furthermore, as pointed out by Ref. \cite{hubener_engineering_2021}, the structure of the Hilbert space of the cavity-embedded system is extremely similar to that of a classically-driven system from the point of view of the Floquet extended space \cite{eckardt_high-frequency_2015}. Indeed, the Hamiltonian can be written as a block-matrix with the blocks corresponding to the number of photon (real or ``Floquet") sectors. However, key differences subsist. Namely, for the cavity the indices $m$ are positive integers while for Floquet theory $m$ runs over all integers. Furthermore, the operators creating $m$ Floquet photons $\sigma_m$ commute with one another while for real photons $\left[a, a^{\dagger}\right] \neq 0 $, leading to interactions in the effective Hamiltonian for the electrons. Or equivalently, the off-diagonal blocks are not of Toeplitz form for the cavity-matter Hamiltonian. 

In the present subsection, a High Frequency Expansion for the effective Hamiltonian is derived without the need for Floquet theory through a unitary transformation, showing identical results.

\subsection{Equivalence between the Floquet HFE and Van Vleck perturbation theory}\label{subsec:EquivalenFloquetVanVleck}

Similarly the procedure of Ref.~\cite{eckardt_high-frequency_2015} deriving the Floquet HFE in extended space, Van Vleck degenerate perturbation theory is used to perturbatively block-diagonalize the Hamiltonian with respect to the photon number sectors. 
Formally, it is similar to a Schrieffer-Wolff transformation, as it relies on a unitary transformation to derive a perturbative theory. In our particular case, the similarity extend further as we want to describe the physics of a low-energy (compared to $\omega_c$) sector of the Hilbert space which incorporates the virtual processes through the high-energy sector. The similarity between Schrieffer-Wolf transformation and Floquet High Frequency expansion has been reported in the literature \cite{bukov_schrieffer-wolff_2016}.

In the present case, Van Vleck perturbation theory is formalized by the action of a unitary transformation upon the Hamiltonian of the light-matter system : 
\begin{equation} \label{eq:DefinitionVanVleckUnitary}
\hat{U}_{vV} = \text{exp}\left(\sum_{\ell}  \hat{G}^{(\ell)}\right),
\end{equation}
with $ \hat{G}^{(\ell)}$ an anti-Hermitian operator which is of order $\delta^{\ell}$ with $\delta = t/\omega_c$ is the small parameter of our perturbation theory given by the electrons' energy scale relative to the cavity frequency. 

The generators $\hat{G}^{(\ell)}$ are taken to be off-diagonal in the photon number (ansatz), and as shown in App.~\ref{subsec:appFloquetVanVleck} can be chosen so as to make the rotated Hamiltonian photon conserving up to a correction of order $\delta^{n +1}$ :
\begin{equation*}\label{eq:UnitaryTransformationConservation}
    \left.{\hat{U}_{vV}^{[n]} }\right.^{\dagger}\hat{H}\hat{U}_{vV}^{[n]} = \hat{H}_{eff}^{[n]} + \mathcal{O}(\delta^{n +1}) \text{ with } \commut{\hat{H}_{eff}^{[n]} }{\hat{n}} = 0,
\end{equation*}
where $\hat{U}_{vV}^{[n]} = \text{exp} \sum_{\ell = 1}^{n}  \hat{G}^{(\ell)}  $ is the truncated Van Vleck unitary transformation. Moreover, the expression for $\hat{G}^{(\ell)}$ can be found order by order as per App.~\ref{subsec:appFloquetVanVleck}.

Furthermore, the effective Hamiltonian and the Floquet Hamiltonian appear to be related through :
\begin{equation*}% \label{eq:RelationVanVleckFloquet}
    \hat{H}_{eff}^{[n]} = \hbar\omega_c a^{\dagger}a + \hat{H}_{F}^{[n]}
\end{equation*}  
This way, we have shown that the effective Hamiltonian obtained through the Floquet HFE in the rotating frame coincides with that derived through Van Vleck perturbation theory on the High Frequency cavity, with care been taken to go back in the non-rotating frame to add the $\hbar\omega_c \hat{n}$ contribution. 

The unitary transformation ${\hat{U}_{vV}^{[n]} }$, which corresponds to the micromotion within Floquet theory, mixes the light and matter degrees of freedom so that the photons are (perturbatively) conserved. Once the Hamiltonian is photon-conserving, the projection onto the zero-photon sector is exact. Hence, Van Vleck perturbation theory allows us to trade a theory which would be perturbative in the light-matter coupling strength $g$ for a perturbative expansion controlled by $\delta = t/\omega_c$.

A remark can be made that, at order $n$ in the HFE, the effective Hamiltonian is contains terms made up of a product of $2(n+1)$ fermionic operators. Thus, as we get closer to the resonant regime, the truncation of the Floquet Hamiltonian fails and Hamiltonian becomes highly non-local in the sense that it is made up of terms which are a product of an extensive number of fermionic operator.

\subsection{Link to the quantum Floquet picture}

In Ref.~\cite{li_effective_2022}, a High Frequency Expansion for a tight-binding model coupled to a multimode cavity is derived through Brioullin-Wigner perturbation theory. It is done within the quantum Floquet picture where the operators on the tensor product space are decomposed as : 
\begin{equation}\label{eq:QuantumFloquetHamiltonian}
    \hat{H} = \sum_{\mathbf{n},\mathbf{m}} \hat{H}_{\mathbf{n},\mathbf{m}} \otimes \ket{\mathbf{n}}\bra{\mathbf{m}},
\end{equation}
with  $\hat{H}_{\mathbf{n},\mathbf{m}} \equiv \bra{\mathbf{n}}\hat{H}\ket{\mathbf{m}}$ is an operator acting on the electrons Hilbert space $\mathcal{H}_{el}$. In App.~\ref{subsec:appQuantumFloquet}, it is shown that the photon-conserving effective Hamiltonian $\hat{H}_{eff} = \sum_{\mathbf{n}} \hat{H}_{eff,\mathbf{n}} \otimes \ket{\mathbf{n}}\bra{\mathbf{m}}$ with :
\begin{equation} \label{eq:MultimodeEffectiveHamiltonian}
    \hat{H}_{eff,\mathbf{n}} = \hat{H}_{\mathbf{n},\mathbf{n}} + \sum_{\mathbf{m}\neq \mathbf{n}} \frac{ \hat{H}_{\mathbf{n},\mathbf{m}} \hat{H}_{\mathbf{m},\mathbf{n}}   }{(\mathbf{n}-\mathbf{m})\cdot \mathbf{\omega}}+ \mathcal{O}(\frac{1}{\omega_2}),
\end{equation}
can be derived through Van Vleck perturbation theory. This procedure is more usual than the use of Brillouin-Wigner perturbation theory and also gives the unitary transformation which allows to block-diagonalize the Quantum Floquet Hamiltonian of Eq.~(\ref{eq:QuantumFloquetHamiltonian}). As we will see in the next sub-section as well as in the calculation of the squeezing of Sect. \ref{sec:SSH_equilibrium}, this unitary transformation is physically relevant when computing average values of operators in the initial frame.

Furthermore, one shows that Eq.~(\ref{eq:MultimodeEffectiveHamiltonian}) reduces to that of Eq.~(\ref{eq:formulaFloquetHFEatFirstOrder}) in the case of a single mode cavity. 
Even though this article explores in details the physics described by such effective Hamiltonians, it does so for multimode cavities. Coupling to single mode cavity leads to different physics described by the effective Hamiltonian.

\subsection{Importance of the unitary transformation : example of the light-matter entanglement}

In the previous subsections we have used a unitary transformation which mixes light and matter in a non-trivial way to solve our problem, similarly to Ref. \cite{passetti_cavity_2023} and Ref.~\cite{eckhardt_quantum_2022}. When computing expectation of observables, the effect of this unitary transformation has to be taken into account. In the present sub-section the case of the light-matter entenglement is investigated, while a second example of the effect of the unitary transformation is present in the calculation of the squeezing of Sect. \ref{sec:SSH_equilibrium}

The effective Hamiltonian derived within the HFE is photon-conserving. As such, its eigenstates are product states of an electronic state and a photonic Fock state. Thus, in the frame obtained by a unitary transformation, the ground state is always factorizable and the light-matter entanglement is zero. However, taking into account the effect of the unitary transformation leads to light-matter entanglement in the initial basis of the problem. This is due to the fact that $\hat{U}^{[\ell]} \neq \hat{U}_{el}\otimes\hat{U}_{ph}$, so that the partial trace on the photon sector is not trivial.

Starting from the Van Vleck frame ground state $\left|\Psi_0\right> = \left|0\right>_{ph}\otimes \left|\psi_0\right>_{el}$, the density matrix of the light-matter system is : 
\begin{equation} \label{eq:densityMatrixElectronPhoton}
\hat{\rho}_0 = \left| 0 \right>\left< 0 \right|\otimes\left| \psi_0 \right>\left< \psi_0 \right|.
\end{equation}
The reduced matrix for the electrons in the initial frame is defined as :
\begin{equation*} \label{eq:densityMatrixElectron_definition}
    \hat{\rho}_{el} = \text{tr}_{ph}\left( \hat{U}_{vV}^{\dagger} \hat{\rho}_0\hat{U}_{vV} \right),
\end{equation*}
which we expand to second order in $\frac{1}{\omega_c}$ by writing the unitary transformation $\hat{U}_{vV}= e^{\hat{G}^{(1)} + \hat{G}^{(2)}}$ where $\hat{G}^{(1)}$ and $\hat{G}^{(2)}$ are anti-hermitian operators whose form are known from Van Vleck perturbation theory and in particular, $\hat{G}^{(1)}$ is given in Appendix~\ref{sec:AppendixVanVleck}. The unperturbed reduced electronic density matrix is defined as $\hat{\rho}_{el,0} = \text{tr}_{ph}\left( \rho_0 \right) = \left| \psi_0 \right>\left< \psi_0 \right|  $. To second order, the electronic reduced density matrix reads : 
\begin{align*} 
    \hat{\rho}_{el} &= \text{tr}_{ph}\left( e^{-\hat{G}^{(1)} - \hat{G}^{(2)}}\hat{\rho}_0e^{\hat{G}^{(1)} + \hat{G}^{(2)}} \right)\\
    &= \text{tr}_{ph}\left( \rho_0 \right) - \text{tr}_{ph}\left(\left[\hat{G}^{(1)}, \rho_0\right] \right) - \text{tr}_{ph}\left(\left[\hat{G}^{(2)}, \rho_0\right] \right) \\&+\frac{1}{2} \text{tr}_{ph}\left(\left[\hat{G}^{(1)}, \left[\hat{G}^{(1)}, \rho_0\right]\right] \right),
\end{align*}
by definition $\hat{G}^{(1)}$ and $\hat{G}^{(2)}$ are block-off diagonal in the photon number, while $\hat{\rho}_0$ is clearly diagonal in the photon number. As such, the second and third term vanish. Then, %$\text{tr}_{ph}\left(\left[\hat{G}^{(1)}, \rho_0\right] \right) = \text{tr}_{ph}\left(\left[\hat{G}^{(2)}, \rho_0\right] \right) =0$
\begin{align*} 
    \hat{\rho}_{el} &= \hat{\rho}_{el,0} + \frac{1}{2}\text{tr}_{ph}\left(\left[\hat{G}^{(1)}, \left[\hat{G}^{(1)}, \rho_0\right]\right] \right)\\
% &= \hat{\rho}_{el,0} + \frac{1}{2}\text{tr}_{ph}\left(\left(\hat{G}^{(1)}\right)^2\rho_0 + \rho_0 \left(\hat{G}^{(1)}\right)^2 - 2 \hat{G}^{(1)}\rho_0\hat{G}^{(1)}\right)\\
&= \hat{\rho}_{el,0} + \frac{1}{2} \left\{ \left<0\right| \left(\hat{G}^{(1)}\right)^2\left|0\right>, \hat{\rho}_{el,0}  \right\}\\& - \sum_{n\in\N} \left<n\right|  \hat{G}^{(1)}\left|0\right>  \hat{\rho}_{el,0} \left<0\right|  \hat{G}^{(1)}\left|n\right> .
\end{align*}
From Eq.~(\ref{eq:FourierCoeffTB_Gamma}), the operators $\left<0\right|  \hat{G}^{(1)}\left|n\right>$,  $\left<n\right|  \hat{G}^{(1)}\left|0\right> $ and $\left<0\right| \left(\hat{G}^{(1)}\right)^2\left|0\right>$ acting on the electrons Hilbert space can be computed and found to be given by 
\begin{widetext}
    % \begin{subequations} 
        \begin{align*}
            \left<0\right| \left(\hat{G}^{(1)}\right)^2\left|0\right> &= -\sum_{ijkl}\sum_{n>0}\frac{t_{i,j}t_{k,l}}{n^2\omega_c^2} \left< 0\right| \hat{\gamma}^{-n}(g_{i,j})\left|n\right>\left<n\right|\hat{\gamma}^{n}(g_{k,l})\left|0\right> c_i^{\dagger}c_jc_k^{\dagger}c_l  ,\\
            \left<n\right|  \hat{G}^{(1)}\left|0\right> &= - \left(\left<0\right|  \hat{G}^{(1)}\left|n\right> \right)^{\dagger}  = \sum_{ij}\frac{t_{ij}}{n\omega_c}\left<n\right|\hat{\gamma}^{n}(g_{k,l})\left|0\right> c_i^{\dagger}c_j \text{ for }n>0 ,%\\
            % \left<0\right|  \hat{G}^{(1)}\left|n\right> &= -\sum_{ij}\frac{t_{ij}}{n\omega_c}\left<0\right|\hat{\gamma}^{-n}(g_{k,l})\left|n\right> c_i^{\dagger}c_j \text{ for }n>0,
        \end{align*}
    % \end{subequations}
\end{widetext}
where it has been used that $\hat{\gamma}^{n}$ creates $n$ photon (algebraically) so as to insert $\left|n\right>\left<n\right|$ in the first equation.
% Then, the reduced electronic density matrix can be written :
% \begin{align*} 
%     \hat{\rho}_{el} &= \hat{\rho}_{el,0} - \frac{1}{2} \sum_{n>0}\sum_{ijkl}\frac{t_{i,j}t_{k,l}}{n^2\omega_c^2}\left< 0\right| \hat{\gamma}^{-n}(g_{i,j})\left|n\right>\left<n\right|\hat{\gamma}^{n}(g_{k,l})\left|0\right> \left(   \left\{c_i^{\dagger}c_jc_k^{\dagger}c_l  , \hat{\rho}_{el,0}\right\}  -  2 c_i^{\dagger}c_j\hat{\rho}_{el,0}c_k^{\dagger}c_l \right).
% \end{align*}
Then, one can define the following operators acting on the electrons  
\begin{equation} \label{eq:Nphoton_PseudoCurrent}
\hat{\mathcal{J}}_n = \sum_{ij}\frac{t_{i,j}}{n\omega_c}\left< 0\right| \hat{\gamma}^{-n}(g_{i,j})\left|n\right>c_i^{\dagger}c_j,
\end{equation}
and using the symmetry of the gamma function $\left(\hat{\gamma}^{m}(g)\right)^{\dagger} = \hat{\gamma}^{-m}(-g)$ as well as $t_{i,j}^{\ast} = t_{j,i}$ and $g_{i,j} = - g_{j,i}$%, the complex conjugate of $\hat{\mathcal{J}}_n $ is given by :
% \begin{align*} 
%     \left(\hat{\mathcal{J}}_n \right)^{\dagger} &= \sum_{ij}\frac{t_{i,j}}{n\omega_c}\left< n\right| \hat{\gamma}^{n}(g_{i,j})\left|0\right>c_i^{\dagger}c_j,
% \end{align*}
% so that
, one shows that the reduced electronic density matrix reads at the leading order in $\omega_c$ : 
\begin{equation} \label{eq:reducedDensityMatrix_Jn}
    \hat{\rho}_{el} = \hat{\rho}_{el,0} - \frac{1}{2} \sum_{n>0} \left(   \left\{\hat{\mathcal{J}}_n \hat{\mathcal{J}}_n^{\dagger}  , \hat{\rho}_{el,0}\right\}  -  2 \hat{\mathcal{J}}_n \hat{\rho}_{el,0}\hat{\mathcal{J}}_n^{\dagger} \right).
\end{equation} 
% As one may expect, the reduced density matrix fulfills an equation similar to that of the Linblad Master equation with the jump operators $\hat{\mathcal{J}}_n$.

Then, the Renyi entropy $S_2(\hat{\rho}_{el}) = - 1/2 \ \text{ln}\ \text{tr} \hat{\rho}_{el}^2  $ reads as : 
\begin{equation} \label{eq:EntenglementEntropyFluctuation}
    S_2(\hat{\rho}_{el}) =  \sum_{n>0}  \left( \left<\hat{\mathcal{J}}_n \hat{\mathcal{J}}_n^{\dagger}  \right> - \left<\hat{\mathcal{J}}_n\right>\left< \hat{\mathcal{J}}_n^{\dagger}  \right> \right).
\end{equation}
Similarly to the result of Ref.~\cite{passetti_cavity_2023}, the cavity and matter are entangled through the fluctuations of an electronic operator.

\section{High-Frequency Expansion for A Driven Cavity}\label{sec:DrivenHFE}

In this Section we extend the high-frequency expansion to the case in which the electron-cavity problem is classically driven. This is natural within the framework of Floquet theory discussed so far, as long as the classical drive is resonant with the cavity mode and the highest scale in the problem. We discuss both the case in which the cavity is driven and the case in which the electronic system is driven, and obtain the resulting effective Hamiltonian and associated dynamics.

We consider a non-harmonic but $T = \frac{2}{\pi}$-periodic driving with zero average value. The time-dependent Hamiltonian of the electron-photon problem reads
\begin{align}% \label{eq:}
    \hat{\mathcal{H}}(t) &=  \sum_{i,j} t_{i,j} e^{-\ii(g_{i,j}a^{\dagger} + g_{i,j}^{\ast} a  ) } c_i^{\dagger}c_j + \hat{H}_{int} + \omega_c a^{\dagger} a \notag\\
    &+ \sum_{m>0} \eta_{m} e^{-\ii m \omega_{c} t} a^{\dagger} + \eta_m^{\ast} e^{\ii m \omega_{c} t} a
\end{align}
Performing the same unitary transformation $\hat{W}(t)$ as in Sec.~\ref{sec:HFEgeneralprogram} to go in a frame rotating at $\omega_c$ leads to  
$\hat{H}(t) =\hat{W}^{\dagger} \hat{\mathcal{H}(t)}\hat{W} + i \left(\partial_t \hat{W}^{\dagger}\right)W$ which takes the form 
\begin{align}% \label{eq:}
    \hat{H}(t) &=  \sum_{i,j} t_{i,j} e^{-\ii(g_{i,j}e^{\ii\omega_c t }a^{\dagger} + g_{i,j}^{\ast}e^{-\ii\omega_c t } a  ) } c_i^{\dagger}c_j + \hat{H}_{int} \notag \\
    &+ \sum_{m\geq0} \eta_{m+1} e^{-\ii m \omega_{c} t} a^{\dagger} + \eta_{m+1}^{\ast} e^{\ii m \omega_{c} t} a
\end{align}
Expanding the Hamiltonian in harmonics of the drive frequency give the Fourier coefficients
\begin{equation} 
    \label{eq:FourierCoefficientsDriven}
    \hat{H}_m = \hat{H}_m[\eta = 0] + \left\{\begin{array}{c} \eta_{m+1}^{\ast} a \text{ if } m>0 \\ \eta_{m+1} a^{\dagger} \text{ if } m<0 \\
        \eta_1 a^{\dagger}+ \eta_1^{\ast} a \text{ if } m=0
    \end{array}\right.
\end{equation}
where $\hat{H}[\eta = 0]$ and $\hat{H}_m[\eta = 0]$ are respectively the Hamiltonian of the non-driven system described in Sect. \ref{sec:HFE} and its the Fourier coefficients.

\paragraph{Zeroth-order Floquet Hamiltonian}
From this, the zeroth order Floquet Hamiltonian is given by :
\begin{equation} %\label{eq:}
    \hat{H}_F^{(0)} =  \hat{H}_F^{(0)}[\eta = 0] +\eta_1 a^{\dagger}+ \eta_1^{\ast} a.
\end{equation}
Thus, at zeroth order, the Floquet Hamiltonian is the equilibrium one with an added coherent displacement term making it no longer photon-conserving. While seemingly simple, the physics of the problem are determined by the interplay of the drive and non-linearity in the photon subspaces. 

\paragraph{First-order Floquet Hamiltonian}
From Eq.~(\ref{eq:FourierCoefficientsDriven}) the first order Floquet Hamiltonian in the HFE is given by 
\begin{align*} 
\hat{H}_F^{(1)} &= \sum_{m>0} \frac{\commut{\hat{H}_m}{\hat{H}_{-m} }}{m\omega_c}\\
&{=}  \hat{H}_F^{(1)}[\eta =0] + \sum_{m>0} \eta_{m+1}\frac{\left[\hat{H}_m[\eta=0],a^{\dagger} \right]}{m\omega_c}\\& + \sum_{m>0}\eta_{m+1}^{\ast}  \frac{\left[a,\hat{H}_{-m}[\eta=0] \right]}{m\omega_c}.
\end{align*}
As we see, the first order Floquet Hamiltonian involves now commutators of the photon operator $a$ and the $\hat{\gamma}_m$ operators defined in Eq.~(\ref{eq:DefGammaTB}). These commutators can be computed using Eq.~(\ref{eq:FourierCoeffTB_Gamma}), leading to the following result for the first order Floquet Hamiltonian
\begin{widetext}
\begin{equation} \label{eq:FirstOrderHamiltonian_DrivingTheCavity}
    \hat{H}_F^{(1)} =  \hat{H}_F^{(1)}[\eta =0] +\sum_{i,j} t_{i,j} \sum_{m>0}\frac{\ii}{m\omega_c} \parent{\eta_{m+1}^{\ast} g_{i,j}\hat{\gamma}_{-(m+1)}(g_{i,j})  +  \eta_{m+1} g_{i,j}^{\ast} \hat{\gamma}_{m+1}(g_{i,j}) } \otimes c_i^{\dagger}c_j
\end{equation}
\end{widetext}
Again, this effective Floquet Hamiltonian is not photon conserving because $\hat{\gamma}_{m}\left(g_{i,j}\right)$ creates algebraically $m$ photons. Furthermore, we can see that at first order the classical drive only generates new hopping terms for the electrons, coupled to $m$ photon transitions controlled by the $m$-th harmonic of the drive. Overall we see that with this approach we have traded the solution of a periodically driven electron-photon problem with a static one where however photon number is not conserved but evolves non-trivially due to the static drive.

% \textcolor{green}{Should I introduce a model with a simple chain and driving at 2 $\omega_c$ to see the current to squeezing term.}

\subsection{Driving the electrons}
In this subsection we consider instead the case of the electrons being embedded in a cavity and shined on by a laser, which corresponds to a different way of driving out of equilibrium the electron-photon system. The coupling of the electrons to the classical light is done through the Peierls phase, so that the Hamiltonian for the system considered is :    
\begin{equation} \label{eq:initialHamiltonian_CoherentDrivingOfTheElectrons}
    \hat{\mathcal{H}}(t) =  \sum_{i,j} t_{i,j} e^{-\ii(g_{i,j}a^{\dagger} + g_{i,j}^{\ast} a + \eta_{i,j}e^{\ii\omega_c t} + \eta_{i,j}^{\ast}e^{-\ii\omega_c t} ) } c_i^{\dagger}c_j   + \omega_c a^{\dagger} a,
\end{equation}
where $\eta_{i,j}$ are the Peierls phase associated to the laser light between sites $i$ and $j$. 
Going to the rotating frame, the Hamiltonian reads
\begin{equation} \label{eq:RotatingFrameHamiltonian_CoherentDrivingOfTheElectrons}
    \hat{H}(t) =  \sum_{i,j} t_{i,j} e^{-\ii(g_{i,j}a^{\dagger}e^{\ii\omega_c t} + g_{i,j}^{\ast} a e^{-\ii\omega_c t}+ \eta_{i,j}e^{\ii\omega_c t} + \eta_{i,j}^{\ast}e^{-\ii\omega_c t} ) } c_i^{\dagger}c_j.
\end{equation}
We now distinguish two cases depending on the spatial dependence of the classical drive.

\subsubsection{Driving with a matching laser mode}\label{subsec:drivenMatchingLaser}
In the case where the cavity mode and the laser mode coincide, e.g. uniform, so that one can write : 
\begin{equation} \label{eq:PeierlsPhaseMatching_CoherentDrivingOfTheElectrons}
g_{i,j}\alpha_0 = -\eta_{i,j}.
\end{equation}
In other words, the cavity vector potential $\mathbf{A}_{cav}(\mathbf{r})$ and the laser one $\mathbf{A}_{las}(\mathbf{r})$ have to verify :
\begin{equation} \label{eq:PeierlsPhaseMatching_CoherentDrivingOfTheElectrons_VectorPotentials}
\alpha_0 \times \int_{\mathbf{R}_i}^{\mathbf{R}_j} \mathbf{A}_{cav}(\mathbf{r})\cdot \mathbf{dr} = - \int_{\mathbf{R}_i}^{\mathbf{R}_j} \mathbf{A}_{las}(\mathbf{r})\cdot \mathbf{dr},
\end{equation} 
so that within the long wavelength approximation, the condition is reduced to having coinciding mode profiles for the laser and the cavity. 
% \begin{equation} \label{eq:PeierlsPhaseMatching_CoherentDrivingOfTheElectrons_VectorPotentials_LongWavelength}
%     \alpha_0 \times \mathbf{A}_{cav}(\frac{\mathbf{R}_i + \mathbf{R}_j}{2})\cdot\left(\mathbf{R}_i - \mathbf{R}_j\right)= - \mathbf{A}_{las}(\frac{\mathbf{R}_i + \mathbf{R}_j}{2})\cdot\left(\mathbf{R}_i - \mathbf{R}_j\right),
% \end{equation} 

In this case, going to a displaced frame by using the unitary transformation : 
\begin{align*} 
\hat{D}(\alpha_0)  = e^{\alpha_0 a - \alpha_0^{\ast}a^{\dagger}}
\end{align*}
allows to absorb the laser field into the cavity field, so that the Hamiltonian of Eq.~(\ref{eq:RotatingFrameHamiltonian_CoherentDrivingOfTheElectrons}) becomes in the displaced frame 
\begin{align*} 
    \hat{H}_{\alpha_0}(t) &\equiv\hat{D}(\alpha_0)\hat{H}(t)\hat{D}^{\dagger}(\alpha_0)\\
    &= \sum_{i,j} t_{i,j} e^{-\ii(g_{i,j}a^{\dagger}e^{\ii\omega_c t} + g_{i,j}^{\ast} a e^{-\ii\omega_c t}) } c_i^{\dagger}c_j. %+ \hat{H}_{int}.
\end{align*}
It appears that in this displaced frame, the Hamiltonian is identical to the Hamiltonian in the rotating frame for the undriven case of Eq. (\ref{eq:rotatingTightBindingHamiltonian}). As such, the Floquet Hamiltonian in the displaced rotating frame of the driven problem is the same as that of the equilibrium problem. In particular, it is photon conserving. 

The driving has been absorbed in the state of the cavity, thus the later is changed from the equilibrium situation explored above. Indeed, due to the application of the displacement operator, if the cavity was in its vacuum state before the driving started, the initial state in displaced frame is given by  
\begin{equation} \label{eq:InitialStateDisplacedFrame_CoherentDrivingOfTheElectrons}
\ket{\hat{\Psi}(0)} = \ket{\psi_0}\otimes\ket{\alpha_0},
\end{equation}
while the time Evolution is given by a Floquet Hamiltonian which conserves photon number  
\begin{align}\label{eqn:quench}
    \ket{\hat{\Psi}(t)}  = \sum_{n\geq0} c_n  e^{-\ii\bra{n}\hat{H}_F\ket{n} t } \ket{\psi_0}\otimes \ket{n},
\end{align}
where $\bra{n}\hat{H}_F\ket{n}$ is the Floquet Hamiltonian projected on the $n$-photon sector which is an operator acting on the electrons' Hilbert space and $c_n = \left<n\right|\left. \alpha_0 \right> $. The dynamics in each photon sector, as described by Eq.~(\ref{eqn:quench}) resembles the one of an isolated system evolving after a quantum quench of the parameters, or equivalently can be interpreted as an effective \emph{quantum Rabi picture}, where the system performs (many-body) Rabi oscillations in each photon sector.

Since the Floquet Hamiltonian is photon-conserving at all order, it may seem that the cavity does not heat up from the drive in the sense that when the driving is switched off there remains exactly the same number of photons in the cavity. Indeed, the heating is hidden in the unitary transformation of Van Vleck perturbation theory. As such, the real number of photons measured in the laboratory frame is given by $\hat{n}^{\prime}\equiv \hat{U}_{vV}\hat{n}\hat{U}_{vV}^{\dagger} $. It has a matter component whose expectation value is not conserved. 

From this point of view, the evolution in time of an observable acting on the electrons $\hat{\mathcal{O}}_{el}$ is given by : 
\begin{align} 
\left<\hat{\mathcal{O}}_{el} \right>(t)&=\sum_{n\in\N} |c_n|^2 \bra{\psi_0}e^{\ii\bra{n}\hat{H}_F\ket{n} t }   \hat{\mathcal{O}}_{el} e^{-\ii\bra{n}\hat{H}_F\ket{n} t } \ket{\psi_0} \label{eq:QuantumRabiEvolutionElectronObservable}
%  &= \sum_{m,n\in\N} c_n c_m^{\ast}  \left(\bra{\psi_0}e^{\ii\bra{m}\hat{H}_F\ket{m} t } \right)\otimes \bra{m}  \hat{\mathcal{O}}_{el} \left(e^{-\ii\bra{n}\hat{H}_F\ket{n} t } \ket{\psi_0}\right)\otimes \ket{n} \notag\\
% &=\sum_{m,n\in\N} c_n c_m^{\ast} \left< m\right|\left.n\right>\bra{\psi_0}e^{\ii\bra{m}\hat{H}_F\ket{m} t }   \hat{\mathcal{O}}_{el} e^{-\ii\bra{n}\hat{H}_F\ket{n} t } \ket{\psi_0}\notag\\
\end{align}
It is different to the projection of the Floquet Hamiltonian onto the coherent state $\ket{\alpha_0}$, for which we have seen that we would recover a classical drive in the thermodynamic limit if the classical drive is kept constant in Eq.~\ref{eq:PeierlsPhaseMatching_CoherentDrivingOfTheElectrons}.
An intriguing question which is left for future studies is whether in the thermodynamics limit having a cavity mode changes the heating dynamics of the electrons due to the classical drive.

\subsubsection{Driving with space-dependent light}

Alernatively, one can imagine the opposite situation where the laser light varies differently that the cavity mode, with its amplitude and phase being modulated in space. The hamiltonian of Eq.~\ref{eq:initialHamiltonian_CoherentDrivingOfTheElectrons} can be treated in a manner similar to what is introduced in Ref.~\cite{goldman_periodically_2014}. The driving term in Hamiltonian Eq.~\ref{eq:RotatingFrameHamiltonian_CoherentDrivingOfTheElectrons} is written as the product of the driving by the laser and the "driving" by the cavity, so that the Fourier coefficients appear as the convolution of those of both Peierls phase terms.

This way one obtains the following zeroth and first order Floquet Hamiltonians :
\begin{widetext}
    \begin{equation} \label{eq:ZerothOrderHamiltonian_CoherentDrivingOfTheElectrons_SecondApproach}
        \hat{H}_{F}^{(0)} = \sum_{i,j} \sum_{n\neq0}  t_{i,j}\left[ (-\ii)^n e^{\ii n  \phi_{i,j} }\mathcal{J}_n\parent{ |\eta_{i,j}| }  \hat{\gamma}_{i,j}^{-n} \right] \otimes c_i^{\dagger}c_j + \hat{H}_{int}.
    \end{equation}
    \begin{equation} 
        \hat{H}_{F}^{(1)} = \sum_{i,j,k,l}   \sum_{m\neq0}\frac{t_{i,j}t_{k,l}}{m\omega_c}  \left[ \sum_{n,n^{\prime}\neq 0} (-\ii)^{n+n^{\prime}} e^{\ii (n  \phi_{i,j}+n^{\prime}  \phi_{k,l}) }\mathcal{J}_n\parent{ |\eta_{i,j}| } \mathcal{J}_{n^{\prime}}\parent{ |\eta_{k,l}| }   \hat{\gamma}_{i,j}^{m-n} \hat{\gamma}_{i,j}^{-m-n^{\prime}}   \right]\otimes c_i^{\dagger}c_jc_k^{\dagger}c_l.
    \end{equation} 
\end{widetext}
where $\phi_{i,j} = \text{arg}\ \eta_{i,j}$ is associated to the phase of the field. The result obtained for the effective Floquet Hamiltonian suggests the interesting possibility
that modulating the phase and amplitude of the laser in space could allow some control over the hopping and interaction mediated by the cavity. However, those Floquet Hamiltonian are no longer photon conserving, therefore their solution require to keep track of the full dynamics of the photon. The formulas above could be made simpler by projecting onto the zero photon space. However, this approximation is valid up to a characteristic duration which vanishes in the thermodynamic limit. Indeed, the total power of the drive on the system is given by $E_{laser}^2 \times V$ so the time to create a photon, and for which the empty cavity approximation fail, will scale as $t_{1ph}\sim \frac{\omega_c}{E_{laser}^2 \times V}$. We conclude that the case of a classically driven cavity with space dependent light is an interesting avenue for future research. In the following we will consider a specific application of our formalism, both for static and driven problems, based on a one dimensionl SSH model coupled to a cavity.

%This idea remains to be investigated further. \textcolor{green}{Cite 'laser painted interactions' ?}

\section{Application: SSH chain Coupled to an off-resonant Cavity}\label{sec:SSH}

In the present section we apply our formalism to a SSH chain made up of $2L$ sites coupled to a uniform field cavity as studied in Refs.\cite{dmytruk_controlling_2022,nguyen2024electron,perez-gonzalez_light-matter_2025}. 
The distance between a $A$ and $B$ sites within the same unit cell is given by $b_0$ while the distance between a $A$ and $B$ sites within two neighboring unit cells is given by $1- b_0$. As such, the light-matter Hamiltonian is given by :  
\begin{align}
\hat{H} &= \omega_c a^{\dagger}a +  v \sum_{j = 1}^{L} e^{\ii \frac{g}{\sqrt{L}}b_0 (a^{\dagger} + a)}c_{j,A}^{\dagger}c_{j,B} + h.c.\notag\\
& - w\sum_{j = 1}^{L-1} e^{-\ii \frac{g}{\sqrt{L}}(1-b_0) (a^{\dagger} + a)}c_{j+1,A}^{\dagger}c_{j,B}+ h.c.  \label{eq:SSH_Hamiltonian}
\end{align}
%%%%%%%%%%%%%%%%%%%%%%%%%%%%%%%%%%%%%%%%%%%%%%%%%%%%%%%%%%%%%%%%%%%%%%%%%%%%
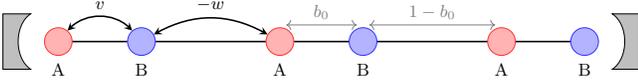
\begin{figure}[t]
    \centering
    \resizebox{0.99\linewidth}{!}{%
        \begin{tikzpicture}[
            every node/.style={font=\small},
            A/.style={circle, draw=red!80, fill=red!30, minimum size=5mm, inner sep=0pt},
            B/.style={circle, draw=blue!80, fill=blue!30, minimum size=5mm, inner sep=0pt},
            arrow/.style={stealth-stealth, thick}
        ]

        % Positions des points
        \node[A] (A1) at (0,0) {};
        \node[B] (B1) at (1.5,0) {};
        \node[A] (A2) at (4,0) {};
        \node[B] (B2) at (5.5,0) {};
        \node[A] (A3) at (8,0) {};
        \node[B] (B3) at (9.5,0) {};

        \node (A2_Prime) at (4,0.3) {};
        \node (B2_Prime) at (5.5,0.3) {};
        \node (A3_Prime) at (8, 0.3) {};

        % Légendes des points
        \node[below=1pt of A1] {A};
        \node[below=1pt of B1] {B};
        \node[below=1pt of A2] {A};
        \node[below=1pt of B2] {B};
        \node[below=1pt of A3] {A};
        \node[below=1pt of B3] {B};

        % Ligne horizontale
        \draw[thick] (A1) -- (B1);
        \draw[thick] (B1) -- (A2);
        \draw[thick] (A2) -- (B2);
        \draw[thick] (B2) -- (A3);
        \draw[thick] (A3) -- (B3);

        % Flèches courbes
        \draw[arrow] (A1) to[bend left=50] node[above] {$v$} (B1);
        \draw[arrow] (A2) to[bend right=30] node[above] {$-w$} (B1);

        % Distances et annotations
        \draw[<->, gray] (A2_Prime) -- node[above] {$b_0$} (B2_Prime);
        \draw[<->, gray] (B2_Prime) -- node[above] {$1 - b_0$} (A3_Prime);
        
        % Grey shapes at the extremities
        % \draw[fill=gray!50, draw=black] (10,0.5) .. controls (9.7,0.3) and (9.7,-0.3) .. (10,-0.5) -- (9.5,-0.5) -- (9.5,0.5) -- cycle;
        % \draw[fill=gray!50, draw=black] (-0.5,0.5) .. controls (-0.2,0.3) and (-0.2,-0.3) .. (-0.5,-0.5) -- (0,-0.5) -- (0,0.5) -- cycle;        
        \draw[fill=gray!50, draw=black] (-0.5,0.5) .. controls (-0.8,0.3) and (-0.8,-0.3) .. (-0.5,-0.5) -- (-1,-0.5) -- (-1,0.5) -- cycle;
        \draw[fill=gray!50, draw=black] (10,0.5) .. controls (10.3,0.3) and (10.3,-0.3) .. (10,-0.5) -- (10.5,-0.5) -- (10.5,0.5) -- cycle;        

        \end{tikzpicture}
    } 
    \centering
    \caption{\justifying Sketch of the SSH chain embedded into a single mode cavity, as described by the Hamiltonian of Eq.~\ref{eq:SSH_Hamiltonian}.}
    \label{fig:DrawingSSH}
\end{figure}
%%%%%%%%%%%%%%%%%%%%%%%%%%%%%%%%%%%%%%%%%%%%%%%%%%%%%%%%%%%%%%%%%%%%%%%%%%%%

To link back this Hamiltonian to the notations of Sect. \ref{sec:HFE}, we can write :
\begin{align*} 
t_{jA,jB} &= t_{jB,jA}^{\ast} = v,\\
t_{j+1A,jB} &= t_{jB,j+1A}^{\ast} = -w,\\
t_{i\alpha, jg} &= 0 \text{ else }. 
\end{align*}
As well as : 
\begin{align*} 
g_{jA,jB} &= -\frac{g}{\sqrt{L}}b_0 ,\\
g_{j+1A,jB} &= \frac{g}{\sqrt{L}}(1-b_0) .
\end{align*}
In the following we will first discuss the equilibrium ground-state properties of the model, focusing in particular on the topological phase diagram in presence of light-matter coupling. Then we will switch to the driven case and discuss the dynamics.

\subsection{Equilibrium}\label{sec:SSH_equilibrium}
\subsubsection{Zeroth order}
\begin{figure}[t]
    \centering
    \includegraphics[width=\linewidth]{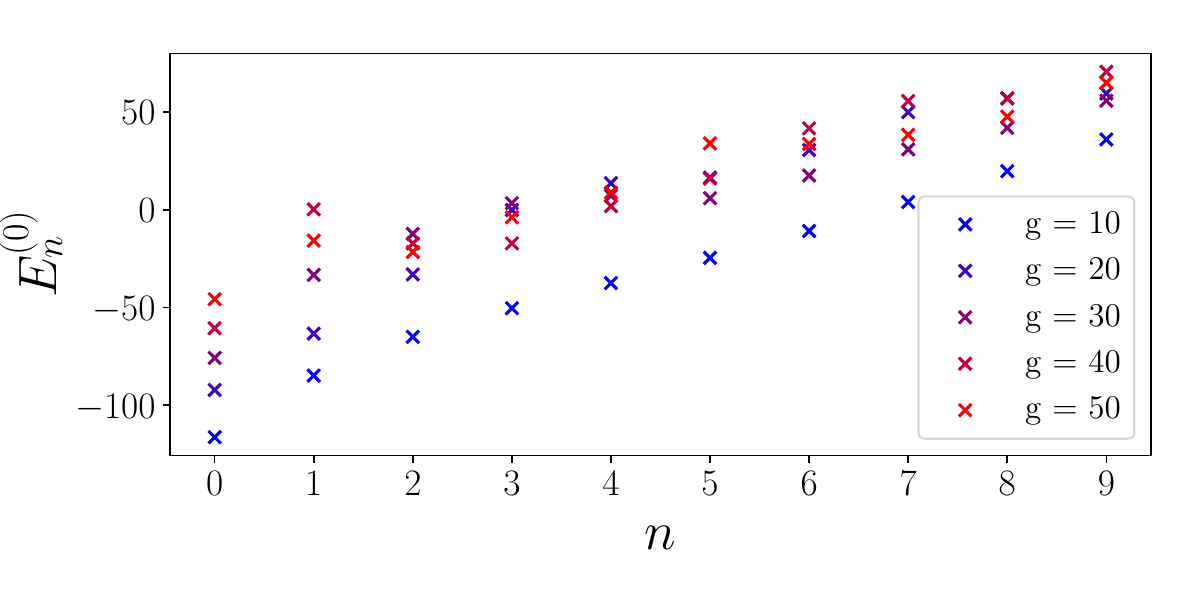}
    \caption{\justifying Ground state energy in the $n$-photon sector for the SSH chain for different values of $g$. The parameters are chosen to be $v =1$, $w = 1.1$, $L = 100$, $b_0 = 0.25$ and $\omega_c = 10$.}
    \label{fig:zerothOrderGroundStateEnergy_nphotonsector}
\end{figure}

Lets us define the following intra- and extra unit-cell hopping global operators :
\begin{subequations} \label{eq:DefTicAndTec}
\begin{align}
    \hat{T}_{i.c.} &\equiv  \sum_{i=1}^L c_{i,A}^{\dagger}c_{i,B},\\
    \hat{T}_{e.c.} &\equiv  \sum_{i=1}^{L-1} c_{i+1,A}^{\dagger}c_{i,B}.
\end{align}
\end{subequations}
%\textcolor{green}{Do I keep those operator ? }
So that, at zeroth order in the HFE, the effective Hamiltonian for the electrons in the $n$-photon sector is given by : 
\begin{equation*}% \label{eq:zerothOrderSSH_nphoton}
    \hat{H}_{F,n}^{(0)} = v \mathcal{I}_{n}\kern-0.25em\left( \frac{gb_0}{\sqrt{L}} \right) \hat{T}_{i.c.}\kern-0.25em - w\mathcal{I}_{n}\kern-0.25em\left( \frac{g(1\kern-0.05em - \kern-0.05em  b_0)}{\sqrt{L}} \right) \hat{T}_{e.c.}\kern-0.25em +h.c.
\end{equation*} 
where $\mathcal{I}_{n}(g) \equiv  e^{-\frac{|g|^2}{2}} \sum_{k} \frac{(-1)^k}{k!}\binom{n}{k} |g|^{2k}$. Hence, in each fixed photon number subspace, the electrons are still described by a SSH model with renormalized coefficients. The energy of the ground state of the electronic system in the $n$ photon sector is easily computed. The value of the cavity+electron system in each subspace at zeroth order, denoted $E_{n}^{(0)}$, is plotted in Fig.~\ref{fig:zerothOrderGroundStateEnergy_nphotonsector}. It shows that in the ground state the cavity is in its vacuum state.
%%%%%%%%%%%%%%%%%%%%%%%%%%%%%%%%%%%%%%%%%%%%%%%%%%
%%%%%%%%%%%%%%%%%%%%%%%%%%%%%%%%%%%%%%%%%%%%%%%%%%
As such, the low-energy electrons are described by the effective SSH Hamiltonian with renormalised parameters
\begin{equation*} 
    %\hat{H}_{F,0}^{(0)} = v e^{-g^2b_0^2/2L} \hat{T}_{i.c.} - we^{-g^2(1 - b_0)^2/2L} \hat{T}_{e.c.} +h.c. 
    \hat{H}_{F,0}^{(0)} = v_{\rm eff}  \hat{T}_{i.c.} - w_{\rm eff} \hat{T}_{e.c.} +h.c. 
\end{equation*} 
where we have defined the following effective hoppings 
\begin{equation} \label{eq:defZerothOrderEffectiveHopping}
    v_{\rm eff} = v e^{-g^2b_0^2/2L} \text{ and }
    w_{\rm eff} = w e^{-g^2(1-b_0)^2/2L} .
\end{equation}
Thus, at zeroth order, the topological phase transition boundary is fixed by the condition $v_{\rm eff}/w_{\rm eff}=1$ as usual for the SSH chain, which reads
\begin{equation} \label{eq:ZerothOrderSSHphaseBoundary}
    \frac{v}{w} e^{\frac{g^2}{2L}(1 - 2 b_0)  } = 1.
\end{equation}
One of the main feature for the ground state found in Ref.~\cite{dmytruk_controlling_2022} is squeezing of the photon. While at first sight it seems absent of our effective description as the photonic ground state is the vacuum, the squeezing is hidden in the unitary transformation used to block-diagonalize the Hamiltonian. What is effectively measured is the operator $\hat{P}^{\prime} = \hat{U}_{vV}\hat{P}\hat{U}_{vV}^{\dagger}$ where $\hat{P} = \frac{\ii}{\sqrt{2}}( a^{\dagger} - a)$ . Using the result of Sect.~\ref{sec:VanVleck},  $\hat{P}^{\prime} $ can be computed at first order in $1/\omega_c$.Then, its fluctuations are found to be equal to : 
\begin{align} \label{eq:FinalResultForSqueezing}
    \left<({\hat{P}^{\prime}})^2\right>_0  &= \frac{1}{2} - \frac{g^2}{2\omega_c L} \sum_k \text{Re}\left( \frac{ v_{\rm eff} - w_{\rm eff} e^{\ii k}}{\left| v_{\rm eff} - w_{\rm eff} e^{\ii k} \right|}\right. \notag \\ &\times(v_{\rm eff}b_0^2 - w_{\rm eff}(1-b_0)^2e^{-\ii k} ) \left. \vphantom{\frac{ v_{\rm eff} - w_{\rm eff} e^{\ii k}}{\left| v_{\rm eff} - w_{\rm eff} e^{\ii k} \right|}}\right)    ,
\end{align}
% \begin{subequations} \label{eq:defZerothOrderEffectiveHopping}
% \begin{align}
%     v_{eff} &= v e^{-g^2b_0^2/2L}  ,\\
%     w_{eff} &= w e^{-g^2(1-b_0)^2/2L} .
% \end{align}
% \end{subequations}
This once again highlights the physical importance of the unitary transformation performed to obtain the effective Hamiltonian thorugh Van Vleck perturbation theory in Sect.~\ref{sec:VanVleck}.
% \textcolor{green}{Plot ? }

\subsubsection{First order}

% \textcolor{green}{I have to add a remark on the fact that when $g$ scales with $\sqrt{L}$, the interactions actually are of order $L^2$. Or from an other point of view, for large system and couplings of the order $g/\sqrt{L}\simeq 1 $ the First order term can be large, interrogating on the convergence of the HFE or the legality of its truncation at high coupling \cite{sentef_quantum_2020} ? Of course everything vanishes in the thermodynamic limit, add a plot on that. }
 
As detailed by Eq.~(\ref{eq:NphotonSector_FirstOrderHamiltonian}), the effective electron-electron interactions appearing in the first order Floquet Hamiltonian are constructed from terms which couple two hopping term initially present in the tight-binding model of Eq.\ref{eq:InitialTightBindingHamiltonian} with a coefficient controlled by the Peierls phase associated to those hopping term. In the present case of the SSH chain, there are 4 inequivalent hopping allowed : from $A$ to be $B$, its opposite, inside a unit-cell or between two unit cells. This gives $4\times 4 = 16$ inequivalent terms in the interactions. % \textcolor{green}{drawing ?}

Projected on the zero photon subspace and Fourier transformed, the Floquet Hamiltonian at first order is given by :
\begin{align}
    &\hat{H}_{el}^{(1)}  = \sum_{k,k^{\prime}} A_{k,k^{\prime}}\ c_{k,A}^{\dagger}c_{k,B} c_{k^{\prime},A}^{\dagger}c_{k^{\prime},B} + h.c. \notag \\
    & + B_{k,k^{\prime}} \left(  c_{k,A}^{\dagger}c_{k,B} c_{k^{\prime},B}^{\dagger}c_{k^{\prime},A} +  c_{k^{\prime},B}^{\dagger}c_{k^{\prime},A} c_{k,A}^{\dagger}c_{k,B}  \right), \label{eq:SSH_firstOrderHamiltonia_inK}
\end{align}
with : 
\begin{subequations} \label{eq:CoefficientsAandB}
\begin{align}
    A_{k,k^{\prime}} &= \frac{v_{\rm eff}^2 }{\omega_c} f(\frac{g^2 b_0^2}{L})+ \frac{w_{\rm eff}^2 }{\omega_c} f(\frac{g^2 (1-b_0)^2}{L}) e^{\ii (k+k^{\prime})} \notag \\&- \frac{v_{\rm eff} w_{\rm eff} }{\omega_c} f\parent{-\frac{g^2}{L}b_0(1- b_0)} (e^{\ii k} +   e^{\ii k^{\prime}} ) ,\\
    B_{k,k^{\prime}} &= \frac{v_{\rm eff}^2 }{\omega_c} f(-\frac{g^2 b_0^2}{L})+ \frac{w_{\rm eff}^2 }{\omega_c} f(-\frac{g^2 (1-b_0)^2}{L}) e^{\ii (k-k^{\prime})}\notag \\& - \frac{v_{\rm eff} w_{\rm eff} }{\omega_c} f\parent{\frac{g^2}{L}b_0(1- b_0)} (e^{\ii k} + e^{-\ii k^{\prime}} ),
\end{align}
\end{subequations}
where $v_{\rm eff}$ and $w_{\rm eff}$ have been defined in Eq.~(\ref{eq:defZerothOrderEffectiveHopping}) and the function $f$ is defined through $f(x)= \int_0^x  \frac{1-e^{-s}}{s}ds$. 
% \begin{widetext}
%     \begin{align} 
%         \hat{H}_{el}^{(1)} &= \frac{v_{eff}^2 }{\omega_c} f(\frac{g^2 b_0^2}{L})  \parent{ \hat{T}_{i.c.}^2 
%         + \parent{\hat{T}_{i.c.}^{\dagger}}^2  } 
%         + \frac{v_{eff}^2 }{\omega_c} f(-\frac{g^2 b_0^2}{L}) \left\{ \hat{T}_{i.c.}, \hat{T}_{i.c.}^{\dagger} \right\} + \frac{w_{eff}^2 }{\omega_c} f(\frac{g^2 (1-b_0)^2}{L})  \parent{ \hat{T}_{e.c.}^2 + \parent{\hat{T}_{e.c.}^{\dagger}}^2  } \notag\\
%         &+ \frac{w_{eff}^2 }{\omega_c} f(-\frac{g^2 (1-b_0)^2}{L}) \left\{ \hat{T}_{e.c.}, \hat{T}_{e.c.}^{\dagger} \right\}
%         - \frac{v_{eff} w_{eff} }{\omega_c} f\parent{\frac{g^2}{L}b_0(1-b_0)} \left( \left\{\hat{T}_{i.c.}, \hat{T}_{e.c.}^{\dagger}\right\} + \left\{\hat{T}_{i.c.}^{\dagger}, \hat{T}_{e.c.}\right\}  \right) \notag\\
%         &- \frac{v_{eff} w_{eff} }{\omega_c} f\parent{\frac{g^2}{L}b_0(b_0-1)} \left( \left\{\hat{T}_{i.c.}^{\dagger}, \hat{T}_{e.c.}^{\dagger}\right\} + \left\{\hat{T}_{i.c.}, \hat{T}_{e.c.}\right\}  \right), \label{eq:FirstOrderSSHhamiltonian_T}
%     \end{align}
% \end{widetext}
% where the operators $\hat{T}_{i.c.}$ and $\hat{T}_{e.c.}$ are those defined in Eq.~\ref{eq:DefTicAndTec}, while $v_{eff}$ and $w_{eff}$ have been defined in Eq.~\ref{eq:defZerothOrderEffectiveHopping}. The function $f$ is defined through $f(x)= \int_0^x  \frac{1-e^{-s}}{s}ds$. 
% Alternate formula : 

Similarly to what is done in Ref. \cite{dag_cavity_2024}, the interactions are treated within Hartree-Fock mean-field decoupling. The interaction Hamiltonian of Eq.~(\ref{eq:SSH_firstOrderHamiltonia_inK}) is thereby rewritten as 
\begin{widetext}
    \begin{align}
        \hat{H}_{el}^{(1)}  &\simeq \sum_{k,k^{\prime}} A_{k,k^{\prime}} \left( \left<c_{k,A}^{\dagger}c_{k,B}\right> c_{k^{\prime},A}^{\dagger}c_{k^{\prime},B} + c_{k,A}^{\dagger}c_{k,B} \left<c_{k^{\prime},A}^{\dagger}c_{k^{\prime},B} \right> -\left< c_{k,A}^{\dagger}c_{k,B} \right>\left<c_{k^{\prime},A}^{\dagger}c_{k^{\prime},B}\right> \right.\notag \\ 
        &+  \left.  \left< c_{k,A}^{\dagger} c_{k^{\prime}, B}  \right>  c_{k,B} c_{k^{\prime},A}^{\dagger} +  c_{k,A}^{\dagger} c_{k^{\prime}, B} \left<c_{k,B} c_{k^{\prime},A}^{\dagger}\right> - \left< c_{k,A}^{\dagger} c_{k^{\prime}, B}  \right>   \left<c_{k,B} c_{k^{\prime},A}^{\dagger}\right> + h.c. \right) + B_{k,k^{\prime}} \dots \label{eq:HartreeFockforSSH}
    \end{align}
\end{widetext}
Since the translation invariance of the system is preserved, it is expected that $\left<\vphantom{c}\right. c_{k,A}^{\dagger} c_{k^{\prime},B}\left. \vphantom{c}\right> \propto \delta_{k,k^{\prime}}$ . Hence, the second line of Eq.~(\ref{eq:HartreeFockforSSH}) is a sum of only $L$ terms, whereas the first line is a sum of $L^2$. Thus, we neglect this ``local" channel of the Hartree-Fock decoupling with respect to the other ``global" channel. 
Following Eq.~\ref{eq:FloquetHamiltonianVacuumCurrentHoppingDecomposition}, the leading order in $g$ for the interaction is of current-current nature. As such, when mean-field decoupled, the global channel term is proportional to expectation value of the current and is negligible because of the absence of macroscopic currents in the ground state which would be equivalent to equilibrium superradiance \cite{andolina_cavity_2019}. As such, the leading contribution to the global channel decoupling of the interactions is controlled by $g^4$, and only appears because the Peierls phase has not been truncated at first order as done for example in Ref. \cite{dag_cavity_2024}. The magnitude of each term in the mean-field Hamiltonian is summarized in the table in Fig.~\ref{fig:tableCoeff}. 
%Contrarily to Ref. \cite{dag_cavity_2024}, we show that at strong coupling and for large system sizes the symmetry breaking term is crushed by the renormalization of the coefficients through the back-action of the electrons through the cavity. \textcolor{red}{this sentence is not clear}

%%%%%%%%%%%%%%%%%%%%%%%%%%%%%%%%%%%%%%%%%%%%%%%%%%%%%%%%%%%%%%%%%%%%%%%%%%%%
\begin{figure}[h]
    \centering
    \resizebox{\linewidth}{!}{
    \begin{tikzpicture}
        % Set column width and row height
        \def\colw{2.5}  % Twice as wide columns
        \def\rowh{1}  % Standard row height
        
        % Draw the table grid
        \draw (0,0) rectangle (\colw*4,\rowh*3); % Outer rectangle
        \foreach \x in {1,2,3} {
            \draw (\x*\colw,0) -- (\x*\colw,\rowh*3); % Vertical lines
    }
    \foreach \y in {1,2} {
        \draw (0,\y*\rowh) -- (\colw*4,\y*\rowh); % Horizontal lines
        }

    \node at ( 0*\colw + \colw/2, 0*\rowh + \rowh/2) {Global channel};
    \node at ( 0*\colw + \colw/2, 1*\rowh + \rowh/2) {Local channnel};  
    \node at ( 0*\colw + \colw/2, 2*\rowh + \rowh/2) {Zeroth order}; 
    \node at ( 1*\colw + \colw/2, 0*\rowh + \rowh/2) {$L\times\frac{g^4}{L^2\omega_c}e^{-g^2/L}$}; 
    \node at ( 1*\colw + \colw/2, 1*\rowh + \rowh/2) {$\frac{g^2}{L\omega_c}e^{-g^2/L}$}; 
    \node at ( 1*\colw + \colw/2, 2*\rowh + \rowh/2) {$e^{-g^2/2L}$}; 
    \node at ( 2*\colw + \colw/2, 0*\rowh + \rowh/2) {$L\times \frac{0.1}{\omega_c}$}; 
    \node at ( 2*\colw + \colw/2, 1*\rowh + \rowh/2) {$\frac{0.1}{\omega_c}$}; 
    \node at ( 2*\colw + \colw/2, 2*\rowh + \rowh/2) {$e^{-g^2/2L}$}; 
    \node at ( 3*\colw + \colw/2, 0*\rowh + \rowh/2) {$L\times \frac{1}{\omega_c} \left(\frac{g}{\sqrt{L}}\right)^{-2} $}; 
    \node at ( 3*\colw + \colw/2, 1*\rowh + \rowh/2) {$\frac{1}{\omega_c}  \left(\frac{g}{\sqrt{L}}\right)^{-2}$}; 
    \node at ( 3*\colw + \colw/2, 2*\rowh + \rowh/2) {$e^{-g^2/2L}$}; 

    % Draw the arrow at the bottom
    \draw[thick, ->] (\colw,0) -- (\colw*4.2,0); 
    % Label at the tip of the arrow
    \node at (\colw*4.1,-0.3) {$\frac{g}{\sqrt{L}}$};
    
    % Labels under the arrow at column demarcations
    \node at (\colw,-0.3) {$0$};
    \node at (\colw*2,-0.3) {$0.5$};
    \node at (\colw*3,-0.3) {$3$};
            
    \end{tikzpicture}
    }
    \centering
    \caption{\justifying Magnitude of the different terms of the mean-field decoupled effective HFE Hamiltonian of the SSH chain. For clarity's sake, $b_0$ has been set to $1/2$ and absorbed in $g$. The $0.1$ prefactor comes from the values of $\mathcal{K}_{0}$ in this region of parameter, as displayed in Fig.~\ref{fig:CoefficientsOddEven}. }
    \label{fig:tableCoeff}
\end{figure}
%%%%%%%%%%%%%%%%%%%%%%%%%%%%%%%%%%%%%%%%%%%%%%%%%%%%%%%%%%%%%%%%%%%%%%%%%%%%

\begin{figure}[ht]
    % First subfigure
    \includegraphics[width=0.45\textwidth]{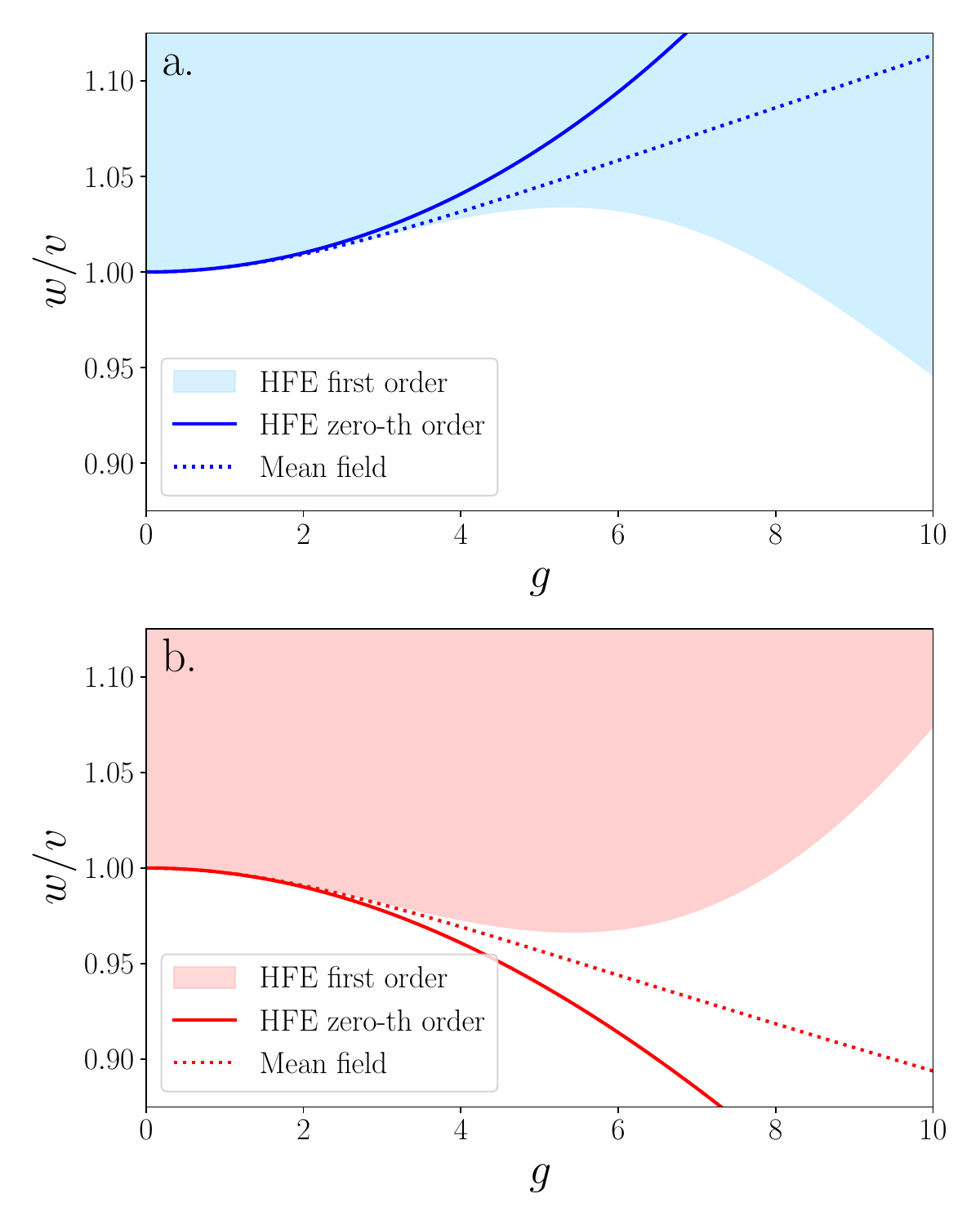}
    % \begin{subfigure}[b]{0.45\textwidth}
    %     \centering        
    %     \includegraphics[width=\textwidth]{Figures/phaseDiagramSSH_B0=25.pdf}
    %     \caption{$b_0 = 0.25$}
    %     \label{subfig:phaseDiagramsSSH_1}
    % \end{subfigure}
    % \hfill
    % % Second subfigure
    % \begin{subfigure}[b]{0.45\textwidth}
    %     \centering
    %     \includegraphics[width=\textwidth]{Figures/phaseDiagramSSH_B0=75.pdf}
    %     \caption{$b_0 = 0.75$}
    %     \label{subfig:phaseDiagramsSSH_2}
    % \end{subfigure}
    
    \caption{\justifying Topological phase diagram of the SSH chain coupled to an off-resonant cavity. The parameters are set to $L = 100$ and $\omega_c = 20$, while in panel a $b_0 = 0.25$ and in panel b $b_0 = 0.75$.
    Depending on the geometric parameter $b_0$ mean-field theory predicts that light-matter coupling $g$ either suppresses ($b_0=0.25$) or enhances ($b_0=0.75$) the topological phase. The HFE results confirm this result at weak coupling $g$, while display a sizable correction to the topological phase diagram at larger values of $g$.}
    \label{fig:phaseDiagramsSSH}
\end{figure}

At first sight, the Hamiltonian of Eq.~(\ref{eq:SSH_firstOrderHamiltonia_inK}) couples sublattice $A$ to sublattice $A$. As such, it seems that the interactions break the chiral symmetry of the SSH model. Within the Hartree-Fock treatment of the interactions the channels breaking the symmetry are controlled by $\left<\vphantom{c}\right. c_{k,A}^{\dagger} c_{k,A}\left. \vphantom{c}\right> $ and $\left<\vphantom{c}\right. c_{k,B}^{\dagger} c_{k,B}\left. \vphantom{c}\right> $. However, the point $\left<\vphantom{c}\right. c_{k,A}^{\dagger} c_{k,A}\left. \vphantom{c}\right>  = \left<\vphantom{c}\right. c_{k,B}^{\dagger} c_{k,B}\left. \vphantom{c}\right>  $ is self-consistent. Therefore we conclude the chiral symmetry is preserved unless the system spontaneously breaks it, which seams unlikely as the term responsible is sub-dominant. Once the mean-field decoupling is performed and the appropriate channels are neglected, one shows that the electrons are once again described by an effective SSH model 
\begin{equation} \label{eq:SSH_modelAtFirstOrderCompleteMeanField}
    \hat{H}_{el}^{[1]} = \mathcal{V}_{eff} \hat{T}_{i.c.} - \mathcal{W}_{eff} \hat{T}_{e.c.} + h.c. ,
\end{equation}
where the coefficients $\mathcal{V}_{eff}$ and $\mathcal{W}_{eff}$ are defined via lengthy expressions reported for completeness in Appendix~\ref{sec:AppendixSSH}. Importantly, these effective hoppings, which arise from decoupling of cavity-mediated interactions, have now to be determined self-consistently by solving for the ground-state of Eq.~(\ref{eq:SSH_modelAtFirstOrderCompleteMeanField}). We now discuss the results obtained by the numerical solution of the self-consistent mean-field equations. Since the electrons under the mean-field approximation are still described by a SSH model, the system can be in topologically non-trivial phase if $\mathcal{W}_{eff}/\mathcal{V}_{eff} >1$. In Fig.~\ref{fig:phaseDiagramsSSH}, the topological phase diagram of the SSH chain is described, with the results given by the mean-field treatment of the electron+photon of Ref.~\cite{dmytruk_controlling_2022} being compared to that of the HFE at zeroth and first order. The result of all three methods coincide at low coupling, but as the coupling increases the zeroth order fails. Then, at even higher coupling the results provided by the HFE first order and the mean-field treatment disagree strongly. It is expected of the mean-field treatment to be inexact at large values of $g/\sqrt{L}$.
%%%%%%%%%%%%%%%%%%%%%%%%%%%%%%%%%%%%%%%%%%%%%%%%%%%%%%%%%%%%%%%%%%%%%%%%%%%%
%%%%%%%%%%%%%%%%%%%%%%%%%%%%%%%%%%%%%%%%%%%%%%%%%%%%%%%%%%%%%%%%%%%%%%%%%%%%

\begin{figure}[t!]
    \centering
    \includegraphics[width=0.45\textwidth]{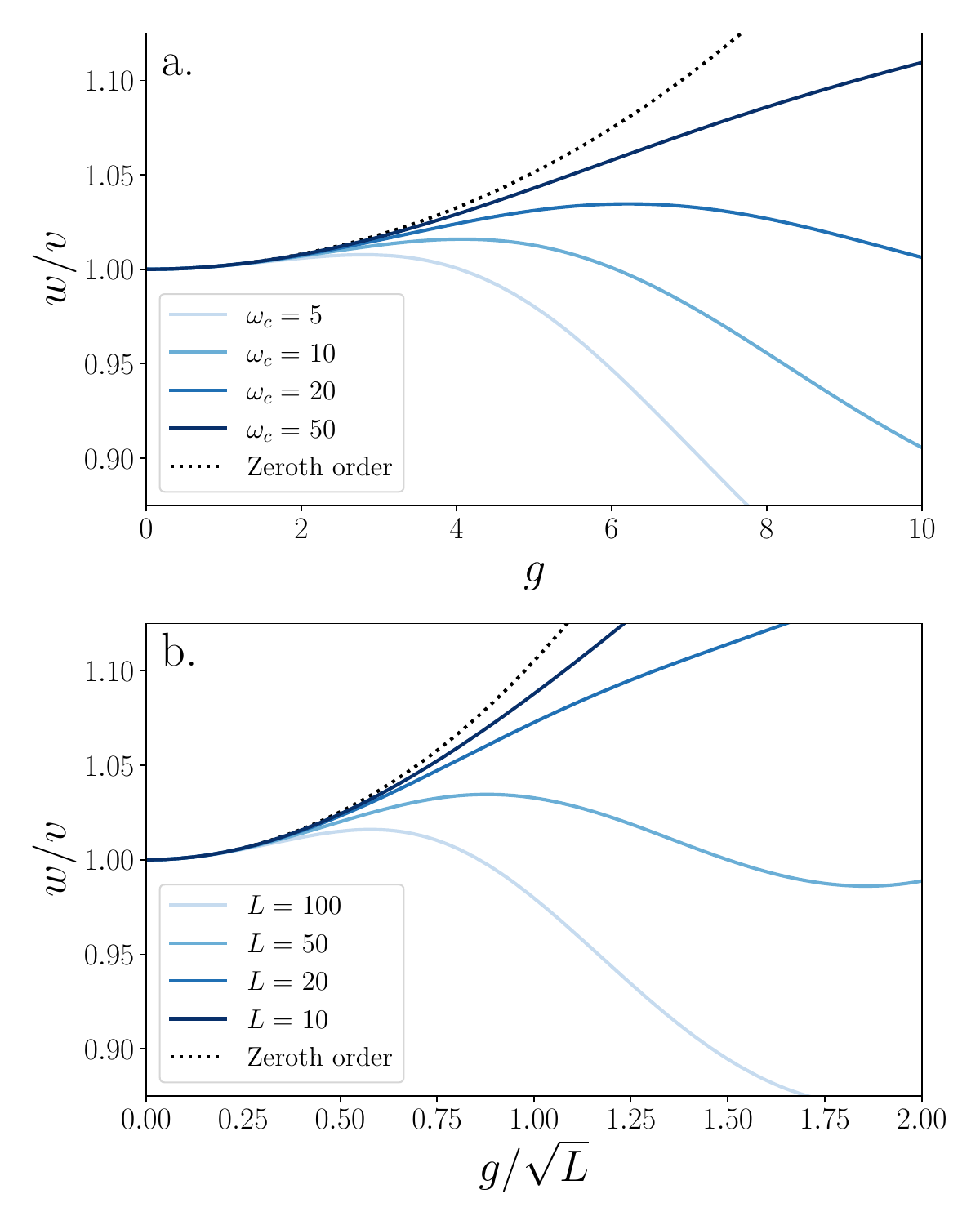}
    % % First subfigure
    % \begin{subfigure}[b]{0.45\textwidth}
    %     \centering        
    %     \includegraphics[width=\textwidth]{Figures/varyingFrequencies.pdf}
    %     \caption{Varying the cavity frequency for $L  = 50$.}
    %     \label{subfig:varyingFrequencies}
    % \end{subfigure}
    % \hfill
    % % Second subfigure
    % \begin{subfigure}[b]{0.45\textwidth}
    %     \centering
    %     \includegraphics[width=\textwidth]{Figures/varyingLengths.pdf}
    %     \caption{Varying the length of the chain for fixed $\omega_c = 20$. It has to be noted that the $g$-axis is rescaled so that the zeroth order result for different lengths to collapse. }
    %     \label{subfig:varyingLengths}
    % \end{subfigure}
    \caption{\justifying Topological phase boundary for the SSH chain coupled to a cavity. Dependence on cavity frequency (panel a) and system size (panel b). The simulations are for $b_0=0.4$. }
    \label{fig:convergenceOftheZerothOrderTruncationHFE}
\end{figure}

Indeed already the analysis of Fig.~\ref{fig:tableCoeff} shows that at a fixed cavity frequency, sufficiently high light-matter coupling lead to a non-negligible contribution of the cavity-mediated interactions to the physics of the system, especially when the system size is large. This also means that truncation of the HFE at zeroth-order like that of Ref.~\cite{sentef_quantum_2020} has to be taken with care at strong light-matter coupling at a fixed cavity frequency. This point is further illustrated in Fig.~\ref{fig:convergenceOftheZerothOrderTruncationHFE}, where we show the comparison between zero-th and first order HFE for the topological phase boundary upon changing the frequency or the system size. We clearly see in panel (a) that sending the frequency to infinity, at a fixed coupling, allows to truncate at zeroth order. However, fixing the frequency and taking large values of $g$ leads to discrepancies between the two results. Similarly, in panel (b) we show that at high coupling with $g/\sqrt{L}$ fixed, the topology of the system depends non-trivially on the size of the system. In particular, the larger $L$ with fixed $g/\sqrt{L}$, the stronger the effect of the interactions making the single-particle approach fail.

%%%%%%%%%%%%%%%%%%%%%%%%%%%%%%%%%%%%%%%%%%%%%%%%%%%%%%%%%%%%%%%%%%%%%%%%%%%%

\begin{figure}[t!]
   \centering
   \includegraphics[width=0.45\textwidth]{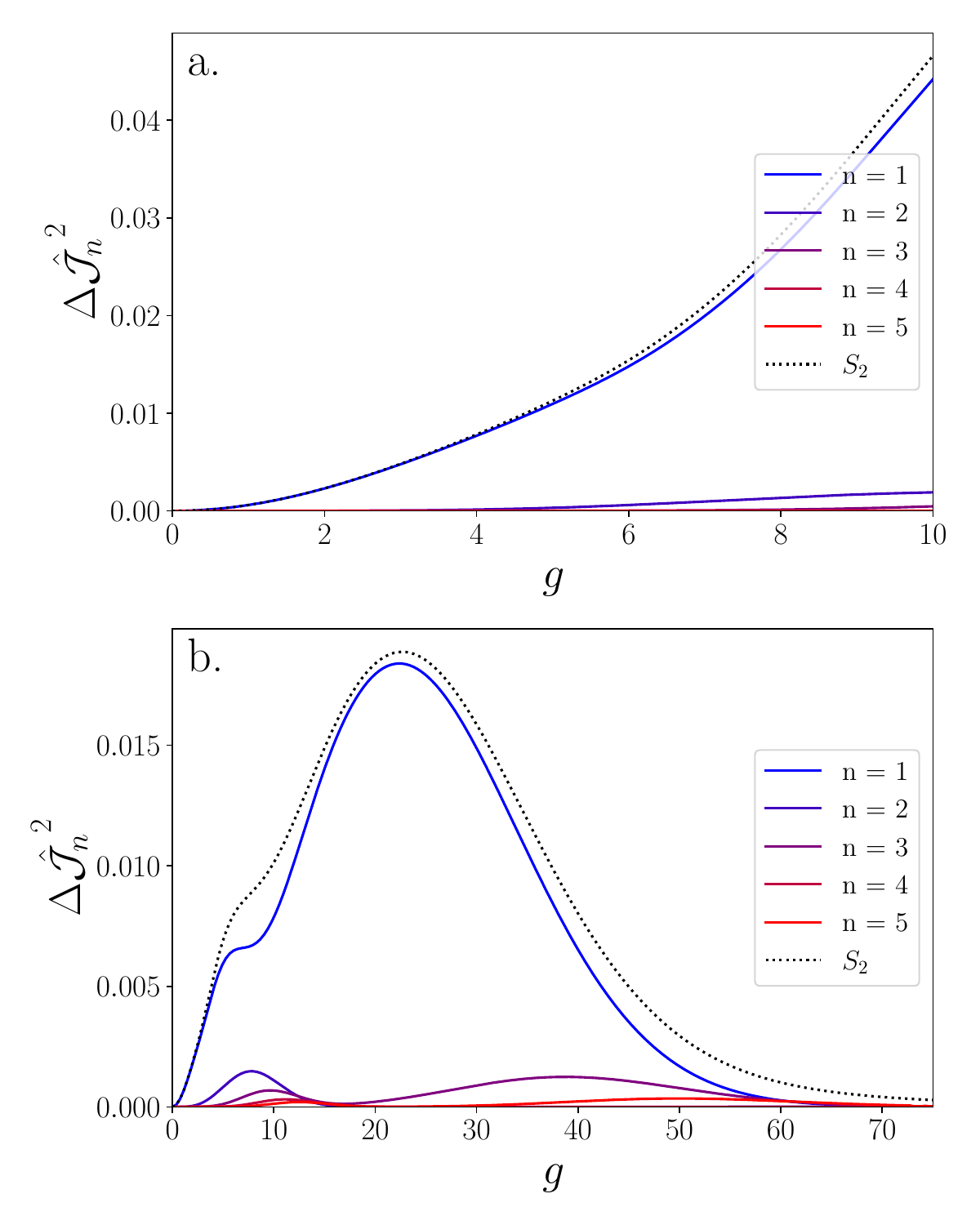}
    % % First subfigure
    % \begin{subfigure}[b]{0.45\textwidth}
    %     \centering        
    %     \includegraphics[width=\textwidth]{Figures/EntropyL=100.pdf}
    %     \caption{$L = 100$}
    %     \label{subfig:entropySSH_1}
    % \end{subfigure}
    % \hfill
    % % Second subfigure
    % \begin{subfigure}[b]{0.45\textwidth}
    %     \centering
    %     \includegraphics[width=\textwidth]{Figures/EntropyL=20.pdf}
    %     \caption{$L = 20$}
    %     \label{subfig:entropySSH_2}
    % \end{subfigure}
    \caption{\justifying Fluctuations of the jump operator $\hat{\mathcal{J}}_n$ and entanglement entropy of the SSH chain. Panel a corresponds to a horizontal cut of Fig.~\ref{fig:phaseDiagramsSSH}a. at $w/v = 1.05$. In panel b, the parameters are set to $v = 1$, $w=1$, $b_0 = 0.25$ and $\omega_c = 20$.}
    \label{fig:entropySSH}
\end{figure}

%DISCUSS FIG 7
We can understand the origin of the disagreement between mean-field results and the HFE by computing the light-matter entanglement entropy of HFE ground state, through the method discussed in Sec.~\ref{sec:VanVleck} (see Eq.~\ref{eq:EntenglementEntropyFluctuation}). This amounts to evaluate the fluctuations of the electronic jump operator $J_n$ which read (see Appendix~\ref{app:CalculationFluctuationsOfJ} for further details)
\begin{equation} \label{eq:formulaFluctuationsOfJ_n}
\left<\hat{\mathcal{J}}_n^2\right> - \left<\hat{\mathcal{J}}_n\right>^2 = \frac{1}{n! n^2}\sum_{k}  \left|j_n(k)\right|^2 \sin^2{\left(\text{arg}j_n(k) - \theta_k\right)  } ,
\end{equation}
where we have defined 
\begin{align} \label{eq:Def_smallJn}
j_n(k) &= \ii^n\left[\frac{v}{\omega_c} e^{-\frac{g^2b_0^2}{2L}} \left( \frac{gb_0}{\sqrt{L}} \right)^n\right.\notag \\
& -\left. \frac{w}{\omega_c} e^{-\frac{g^2(1 - b_0)^2}{2L}} \left( -\frac{g(1-b_0)}{\sqrt{L}} \right)^n e^{\ii k }\right],
\end{align}
as well as $\theta_k \equiv \text{arg}\parent{\mathcal{V}_{eff}  - \mathcal{W}_{eff} e^{-\ii k} }$. The results, shown in Fig.~\ref{fig:entropySSH}, show that as the light-matter coupling $g$ increases, the light-matter entanglement grows~\cite{shaffer_entanglement_2023}. Furthermore it starts growing faster around the value the first order HFE and Mean-Field become qualitatively different (one increase and the other decreases). This seems to indicate that the mean-field fails at this point. Fig.~\ref{fig:entropySSH} shows that the light-matter entanglement is mostly dominated by the entropy produced by the fluctuations of the current associated to the paramagnetic term (corresponding to $n=1$) while the other orders in the Peierls phase only weakly contribute to the light-matter entanglement. They do however generate a slowly decaying tail to the entropy. 
While the diamagnetic terms fluctuation do not entangle light and matter, the diamagnetic term is identified to be the one responsible for the modification of the ground state. 
% \textcolor{green}{I can do a plot where the coupling function $f$ is truncated at 1st order or even just the second order. From the tests I have done, it shows that the diamagnetic term is really the one having a strong effect on the ground state. }

%%%%%%%%%%%%%%%%%%%%%%%%%%%%%%%%%%%%%%%%%%%%%%%%%%%%%%%%%%%%%%%%%%%%%%%%%%%%
%%%%%%%%%%%%%%%%%%%%%%%%%%%%%%%%%%%%%%%%%%%%%%%%%%%%%%%%%%%%%%%%%%%%%%%%%%%%

% \textcolor{green}{Do I talk about the $n$-photon sector problem? Because for the parameters I have chosen in Fig. \ref{fig:phaseDiagramsSSH} it is not actually a problem as n = 0 everywhere.}
% \textcolor{green}{I will add a plot for different $L$ to see that in the thermodynamic limit everything dies out. Shoul I also do $ b_0$ sweep like in Olesia and Marco's article ?  }
To summarize, in this example we have used the HFE developed previously to obtain an effective model of SSH electrons coupled to an off resonant cavity. The model obtained still is of the form of an SSH model, thus the topological nature of the system is easily defined. The HFE results, while in agreement with that of a mean-field treatment for low coupling, show that the mean-field is erroneous and the boundary of the topological transition predicted by it is qualitatively wrong as it is not monotonous. While the original conclusion that the topological properties of a finite-size system can be controlled through a single-mode cavity remains true, the high light-matter couplings required are not described satisfactorily by the light-matter mean-field ansatz.
It is useful to discuss our results in comparison with recent literature. In particular a SSH model coupled to a cavity mode was studied with exact diagonalization in Ref.~\cite{perez-gonzalez_light-matter_2025} (see also the related study in Ref.~\cite{bomantara_quantum-vacuum-protected_2025}, where however a different form of light-matter coupling was considered) which highlighted the breaking of chiral symmetry emerging from the asymmetry of the (quantum) Floquet spectrum. To connect with our results we note that the exact diagonalization result is obtained in the single particle sector, where effects of cavity-mediated interactions are absent (see also discussion in App.~\ref{app:singleElectron}).

\subsection{Driven}

\subsubsection{Driving the cavity at its second harmonic}

This section aims to illustrate the idea developed in Sect.~\ref{sec:DrivenHFE}. We consider the SSH chain embedded in a cavity as described by Eq.~(\ref{eq:SSH_Hamiltonian}). The cavity is coherently driven at $2\omega_c$ with an amplitude $\eta$. In the notations of Sect.~\ref{sec:DrivenHFE} we can write $\eta_m = \eta \delta_{2,m}$. In this particular case, the first order Floquet Hamiltonian from Eq.~(\ref{eq:FirstOrderHamiltonian_DrivingTheCavity}) reads for the SSH chain 
\begin{equation*} %\label{eq:SSH_Hamiltonian_DrivingTheCavity}
\hat{H}_{F}^{(1)} = \hat{H}_{F}^{(1)}[\eta = 0] - \frac{\eta  g^3}{4L\sqrt{L}} (\left. a^{\dagger} \right.^2 + a^2 ) \otimes \hat{\mathcal{J}}_3 + \mathcal{O}\left((\frac{g}{\sqrt{L}})^5  \right),
\end{equation*}
where a generalized current operator $\hat{\mathcal{J}}_3$ has been defined through : 
\begin{equation*} %\label{eq:}
    \hat{\mathcal{J}}_3 = \ii v b_0^3   \left( \hat{T}_{i.c} - \hat{T}_{i.c}^{\dagger} \right)+ \ii w (1-b_0)^3 \left( \hat{T}_{e.c} - \hat{T}_{e.c}^{\dagger}\right).
\end{equation*} 
It can pointed out that in the case where $b_0 = 1/2$, $\hat{\mathcal{J}}_3$ is proportional to the real current operator in the chain. 

We see that the term generated by the driving of the cavity appears as a coupling between the current in the chain and a squeezing term. We emphasize that this term arises because of the interplay between classical drive and light-matter interaction. This Hamiltonian recalls the one of an optical parametric oscillator where an optical non-linearity induces a parametric down-conversion into squeezed states~\cite{collet1984squeezing,gardiner1985input}. We expect that the resulting electron-photon dynamics, where now again photon are not conserved but are enriched of a non-trivial dynamics, will display non-trivial features. The study of this type of problems, involving interacting electrons coupled to driven cavity mode, represents an exciting avenue for future work.

%However, this coupling breaks the photon conservation making the resolution of such equation substantially tougher.
%\textcolor{green}{What to say now ?}
%The study of the dynamics of a such a system is left to be addressed in future work. 

%%%%%%%%%%%%%%%%%%%%%%%%%%%%%%%%%%%%%%%%%%%%%%%%%%%%%%%%%%%%%%%%%%%%%%%%%%%%
\begin{figure}[ht]
    \centering
    \includegraphics[width=0.45\textwidth]{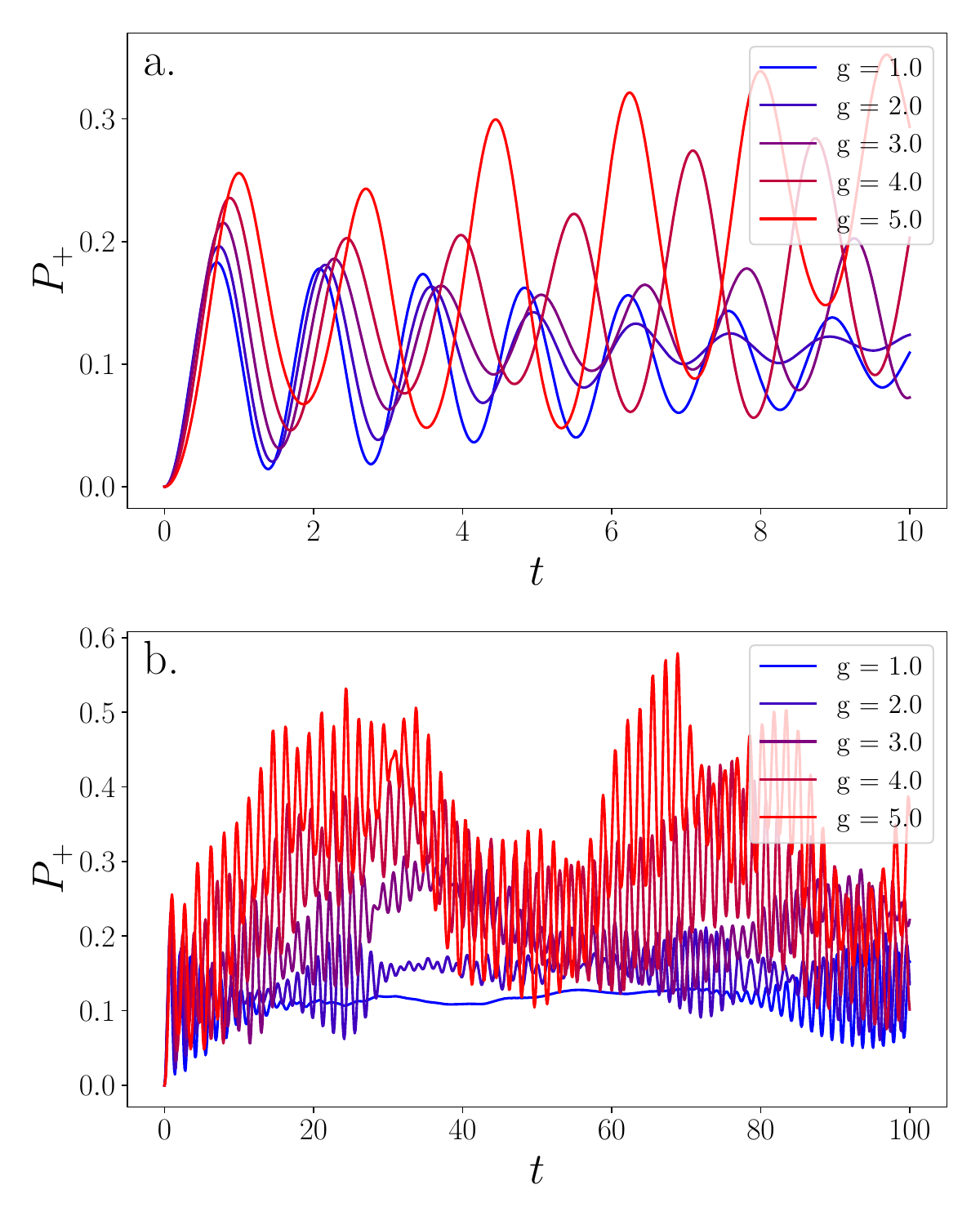}
    % % First subfigure
    % \begin{subfigure}[b]{0.45\textwidth}
    %     \centering        
    %     \includegraphics[width=\textwidth]{Figures/Driven/plotPShortTime.pdf}
    %     \caption{Short time dynamics.}
    %     \label{subfig:DynamicsSSH_1}
    % \end{subfigure}
    % \hfill
    % % Second subfigure
    % \begin{subfigure}[b]{0.45\textwidth}
    %     \centering
    %     \includegraphics[width=\textwidth]{Figures/Driven/plotPLongTime.pdf}
    %     \caption{Long time dynamics.}
    %     \label{subfig:DyanmicsSSH_2}
    % \end{subfigure}
    \caption{\justifying Dynamics of the probability to switch edge state for the driven SSH chain. The parameters are set to $L=20$, $v = 1$, $w = 1.5$, $b_0 = 0.4$, $\omega_c = 10$ and $\eta = 0.5$. The panels respectively display the short and long time dynamics of $P_{+}$ for this set of parameters. }
    \label{fig:dynamicsSSH}
\end{figure}
%%%%%%%%%%%%%%%%%%%%%%%%%%%%%%%%%%%%%%%%%%%%%%%%%%%%%%%%%%%%%%%%%%%%%%%%%%%%

\subsubsection{Driving the electrons}\label{sec:drivenSSH_Rabi}
In this section the formalism developed in Sect.~\ref{subsec:drivenMatchingLaser} is applied to the SSH chain.  Specifically, we consider a SSH chain coupled to a single-mode off-resonant cavity just like in Sect. \ref{sec:SSH}. The assumption is made that the coefficients of the model $v$,$w$,$\omega_c$, $g$, $b_0$ and $L$ are chosen such that the chain is in its topological phase following the results shown in Fig. \ref{fig:phaseDiagramsSSH}. As such, the open boundary chain has end states $\ket{\pm}$ in the gap of the bulk spectrum, with their respective energy $\pm \varepsilon$ which scales as $\varepsilon \propto e^{-\frac{L}{L_0}}$. 

We consider that in the initial state of the system, only $\ket{-}$ is occupied. The system is driven and we are interested in how the occupation of the $\ket{+}$ state evolves in time. The observable of interest is $\P_{+}(t) = \left| \bra{+}\ket{\psi(t)} \right|^2 = \left<  \ket{+}\bra{+} \right>(t)$ . The drive is supposed to match the cavity field and the "quantum Rabi oscillation picture" is used to solve the dynamics. We denote $\eta = \frac{e a}{\hbar}E_{laser}$ the strength of the driving and define $\alpha_0 = -\frac{\eta\sqrt{L}}{g}$ the displacement necessary to absorb the driving field into the cavity. 
Following the discussion in Sec.~\ref{subsec:drivenMatchingLaser} the time evolution of $\hat{P}_{+}$ the projector on $\ket{+}$ reads : 
\begin{equation}\label{eq:dynamicsP} 
    \P_{+}(t) = \sum_{n\geq 0} |c_n|^2 \bra{\psi_0} e^{\ii \hat{H}_{F,n} t}\hat{P}_{+} e^{-\ii \hat{H}_{F,n} t}\ket{\psi_0}
\end{equation}
where $c_n  = e^{-\frac{|\alpha_0|^2}{2}}\frac{\alpha_0^n}{\sqrt{n!}}$ and $\ket{\psi_0}$ is the ground state of the equilibrium system. In Eq.~(\ref{eq:dynamicsP}) the effective Hamiltonian driving the dynamics, $\hat{H}_{F,n}$, is the first order Floquet Hamiltonian in the $n$ photon sector, discussed in generality in Sec.~(\ref{sec:HFE})  and more specifically for the SSH in Sec.~(\ref{sec:SSH_equilibrium}). It contains in particular the cavity mediated interactions whose effect on the topological phase boundary has been discussed previously. To solve for its dynamics, needed to evaluate $\P_{+}(t)$, we resort to a time-dependent mean-field decoupling for which the details can be found in App.~\ref{app:timeDependentMF}.

The results of the driven dynamics are shown in Fig.~\ref{fig:dynamicsSSH}. At short times (panel a) we see coherent oscillations with amplitude and frequency which depend weakly on the light-matter coupling $g$.  However, the long time dynamics' behavior depend heavily on $g$. Indeed, for low values of $g$, $\alpha_0$ is quite large which makes it so a lot of values of $n$ take part in the dynamics through Eq.~(\ref{eq:dynamicsP}). All the different frequencies interfere leading to thermalizing dynamics. On the contrary, for strong couplings only a few terms in Eq.~\ref{eq:dynamicsP} and the dynamics display very slow oscillations. Overall, these results demonstrate that the dynamics of a classically driven SSH coupled to a cavity mode can be rather rich and complex and worth to be explored in the future.

% Talk about the time-dependent mean-field but relegate the details to appendix. Show the results I have.

\section{Conclusions}\label{sec:conclusions}

In this work we have developed a general perturbative procedure, based on Floquet theory and the HFE, to derive an effective Hamiltonian for the electrons strongly coupled to an off-resonant cavity. This effective Hamiltonian incorporates the fluctuations of the cavity field leading to cavity-mediated interactions and second nearest-neighbor hoppings, the former appearing directly because of the quantum-ness of the electromagnetic field in the cavity. We have shown that this approach is fully equivalent to Van Vleck perturbation theory and to Brillouin-Wigner perturbaton theory used in the the quantum Floquet picture, which has allowed us to highlight the importance of the unitary transformation used to block-diagonalize the Hamiltonian, equivalent to the Floquet micromotion operator. This is particularly relevant when calculating for example the light-matter entanglement.

Working within the Floquet framework naturally allows to extend our approach to the case in which either the electronic system or the cavity are periodically driven, at frequency resonant with the cavity mode which is the highest energy scale in the problem. We have shown that including a classical drive gives a non-trivial dynamics to the photonic degrees of freedom, which in the ground-state would be otherwise frozen in a given fixed photon number sector. This makes the electron-photon problem still quite non-trivial to solve. We have shown that progress can be obtained for certain types of driving acting on the electronic degrees of freedom only, where the dynamics can be rewritten as a sort of quantum Rabi oscillations in the photonic sector.

We have applied our theoretical framework to study the physics of a SSH chain coupled to single mode cavity, focusing in particular on its topological properties in presence of light-matter coupling. For this model we have derived the effective Floquet Hamiltonian at zero and first order in the HFE, the latter including non-trivial interactions mediated by the cavity. We have solved the resulting many-body problem within mean-field theory, by mapping the problem back to an effective SSH in self-consistently determined staggered hoppings. We have compared the topological phase boundary obtained in this way with the result of the zero order and a direct mean-field decoupling between electrons and photons. This comparison show that at strong light-matter coupling the cavity mediated interactions become relevant and change the topology of the phase diagram. We have understood this difference to light-matter entanglement, which is captured by our framework via the unitary transformation while is missed in mean-field. Finally, we have also considered the case of a driven SSH model coupled a cavity mode, with the drive acting on either the photon or the electrons. In the former case we have shown the effective dynamics give rise to dynamical squeezing mediating by an electronic current. In the latter, we have shown the complex time evolution of the edge mode.

Our work suggests a number directions for future research. A natural avenue is to study the effect that cavity-mediated interactions can have in driving electronic instabilities, notably superconductivity. The case of a driven cavity, where the photon acquires a non-trivial dynamics even in the HFE, or driven electrons where the classical drive can control the strenght of cavity-mediated interactions deserve further work. From a methodological perspective one can imagine going beyond the high-frequency expansion and follow the analogy between cavity modes and Floquet modes to use other methods to derive effective Hamiltonian, such as the flow-equation approach~\cite{wegner_flow_2001,thomson_flow_2021}, where a series of continuous unitaries is performed to effectively diagonalize the problem.

\section{Acknowledgments}

We acknowledge inspiring discussions with Nathan Goldman. G.M.A acknowledge funding from the European Union’s Horizon 2020 research and innovation programme under the Marie Sklodowska-Curie
(Grant Agreement No. 101146870 --  COMPASS).  M.S. acknowledge funding from the European Research Council (ERC) under the European Union's Horizon 2020 research and innovation programme (Grant agreement No. 101002955 -- CONQUER).

\appendix

\section{Proofs for Van Vleck perturbation theory}\label{sec:AppendixVanVleck}

\subsection{Derivation of the effective Hamiltonian through Van Vleck perturbation theory}\label{subsec:appFloquetVanVleck}

Very generally, the light-matter Hamiltonian is written as :
\begin{equation}
    \hat{H} = \hat{V} + \mathbbm{1}\otimes\hat{n}\omega_c ,
\end{equation}
with $\omega_c$ being the dominant energy scale of the Hamiltonian. The perturbation will be controlled by a small dimensionless parameter $\delta\ll1 $ such that $\hat{V} \sim \delta \omega_c$. We denote $\hat{H}_0 = \mathbbm{1}\otimes\hat{n}\omega_c $ the dominant term.

We introduce the following notations : for any operator $\hat{A}$ in $\mathcal{H} = \mathcal{H}_{el}\otimes \mathcal{H}_{ph}$, we decompose it as :
\begin{equation*}
    \hat{A} = \hat{A}_D + \hat{A}_X ,
\end{equation*}
where $\hat{A}_D$ is block diagonal in the photon number while $\hat{A}_X$ is block off-diagonal. If we denote $\hat{P}_n \equiv\mathbbm{1}\otimes\ket{n}\bra{n} $ the projector on the $n$-photon sector, the decomposition above comes from the closure relation $\sum_n \hat{P}_n = \mathbbm{1} $ : 
\begin{align*}
    \hat{A}_D &\equiv \sum_n \hat{P}_n\hat{A}\hat{P}_n\\
    \hat{A}_X &\equiv \sum_{n\neq m } \hat{P}_n\hat{A}\hat{P}_m.
\end{align*}

The Hamiltonian will be block-diagonalized by performing the unitary transform $\hat{U} = e^{\hat{G}}$ where $\hat{G}$ is anti-hermitian. $\hat{G}$ is chosen to be block off-diagonal (ansatz). 
We define $\hat{W} \equiv \hat{U}^{\dagger}\hat{H}\hat{U} $ and require that $\hat{W}_X = \mathcal{O}(\frac{1}{{\omega_c }^{\ell}})$ where $\ell$ is the truncation order.

We then expand both $\hat{W}$ and $\hat{G} = \sum_{\nu} \delta^{\nu} \hat{G}^{(\nu)}$. At the zeroth order in $\delta$ i.e. when $\delta =0$, the initial Hamiltonian is block diagonal so 
\begin{subequations}
    \begin{align*}
        \hat{W}^{(0)} &= \mathbbm{1}\otimes\hat{n}\omega_c \\
        \hat{G}^{(0)} &= 0   .
    \end{align*}
\end{subequations}
Then, we use the formula $\hat{W} = e^{-\hat{G}}\hat{H}e^{\hat{G}} = \exp\left(\parent{-\text{ad}_{\hat{G}}}\right)\hat{H}$ to write $\hat{W}$ up to order two :
\begin{subequations}\label{eq:ExpansionWOrder1and2}
    \begin{align}
        \hat{W}^{(1)} &= \hat{V} - \commut{\hat{G}^{(1)}}{\hat{H}_0},\\
        \hat{W}^{(2)} &= \frac{1}{2} \commut{\hat{G}^{(1)}}{\commut{\hat{G}^{(1)}}{\hat{H}_0}} - \commut{\hat{G}^{(1)}}{\hat{V}} - \commut{\hat{G}^{(2)}}{\hat{H}_0}   .
    \end{align}
\end{subequations}
Enforcing the off-diagonality of $\hat{W}^{(\nu)}$ imposes that $\hat{G}^{(\nu)}$ satisfies: 
\begin{subequations}\label{eq:GcommutatorEquations}
    \begin{align}
        \commut{\hat{G}^{(1)}}{\hat{H}_0} &= \hat{V}_X ,\label{eq:G1commutatorEquation} \\
        \commut{\hat{G}^{(2)}}{\hat{H}_0} &= \frac{1}{2} \commut{\hat{G}^{(1)}}{\commut{\hat{G}^{(1)}}{\hat{H}_0}}_X - \commut{\hat{G}^{(1)}}{\hat{V}}_X  .
    \end{align}
\end{subequations}
Re-injecting the first in the second equation yields the equation on $\hat{G}^{(2)}$ :
\begin{equation*}
    \commut{\hat{G}^{(2)}}{\hat{H}_0} =\commut{\hat{V}_D}{\hat{G}^{(1)}}+ \frac{1}{2}\commut{\hat{V}_X}{\hat{G}^{(1)}}_X ,
\end{equation*}
and thus $\hat{G}^{(2)}$ can be computed once $\hat{G}^{(1)}$ is known. This way, the generators and effective Hamiltonian can be derived order by order. 

Injecting those equations into \ref{eq:ExpansionWOrder1and2} gives : 
\begin{subequations}\label{eq:ExpansionWOrder1and2Deuxieme}
    \begin{align}
        \hat{W}^{(1)} &= \hat{V}_D\\
        \hat{W}^{(2)} &= \commut{\hat{V}_X}{\hat{G}^{(1)}}_D ,
    \end{align}
\end{subequations}
so that we only need to determine $\hat{G}^{(1)}$ from \ref{eq:G1commutatorEquation} to get $\hat{W}^{(2)}$ up to second order in $\delta$ i.e. first order in $\frac{1}{\omega_c}$. 

In \cite{eckardt_high-frequency_2015} this equation is solved through projecting onto Floquet "photon" number states. Here, to link it to the Floquet HFE in the rotating frame, this will be done through the Fourier transform. 
We multiply Eq. \ref{eq:G1commutatorEquation} on the left by $e^{\ii \omega_c \hat{n}t}$ and by $e^{-\ii \omega_c \hat{n}t}$ on the right. Using $\commut{\hat{H}_0}{\hat{n}} = 0 $ gives :
\begin{equation*}
    \commut{\hat{G}^{(1)}(t)}{\hat{H}_0} = \hat{V}_X(t),
\end{equation*}
where $\hat{G}^{(1)}(t)\equiv e^{\ii \omega_c \hat{n}t}\hat{G}^{(1)}e^{-\ii \omega_c \hat{n}t}$ and $\hat{V}_X(t)\equiv e^{\ii \omega_c \hat{n}t}\hat{V}_Xe^{-\ii \omega_c \hat{n}t}$ are $T$-periodic operators and thus can be expended as Fourier coefficients : 
\begin{equation*}
    \sum_{m\neq0}\commut{\hat{G}^{(1)}_m}{\hat{H}_0}e^{\ii m \omega_c t } = \sum_{m\in\Z}\hat{V}_{X,m}e^{\ii m \omega_c t }.
\end{equation*}
So that Fourier coefficient by Fourier coefficient : 
\begin{equation*}
    \forall m\neq 0, \ \commut{\hat{G}^{(1)}_m}{\hat{H}_0} = \hat{V}_{X,m},
\end{equation*}
where the $m= 0$ is made trivial by the block off-diagonal nature of the time dependent objects. 
From this we can remark that for $m\neq0$ then $\hat{V}_{X,m} = \hat{H}_m$.

Finally, the projection operator $\hat{P}_n$ can be applied on both sides to write : 
\begin{align*}
     \hat{P}_{n}\commut{\hat{G}^{(1)}_m}{\hat{H}_0} \hat{P}_{n^{\prime}} &=  \hat{P}_{n}\hat{V}_{X,m} \hat{P}_{n^{\prime}}\\
    &= (n^{\prime} - n) \omega_c \hat{P}_{n}{\hat{G}^{(1)}_m}\hat{P}_{n^{\prime}}.
\end{align*}
It then has to be noted that because of the shape of the unitary transformation we have used to go into the rotating frame, $\hat{V}_{X,m}$ creates (algebraically) $m$ photons. As such, $\hat{P}_{n}\hat{V}_{X,m} \hat{P}_{n^{\prime}} \propto \delta_{n,n^{\prime}+m}$.
Thus, the previous equation becomes :
\begin{equation*}
    \forall n, n^{\prime}, \ -m\omega_c \hat{P}_{n}{\hat{G}^{(1)}_m}\hat{P}_{n^{\prime}} =\hat{P}_{n}\hat{V}_{X,m} \hat{P}_{n^{\prime}}.
\end{equation*}
So that, taking :
\begin{equation}
    \hat{G}^{(1)}_m = -\frac{\hat{H}_m}{m\omega_c},
\end{equation}
and going back to the non-rotating frame, while utilizing that $\hat{H}_m$ only creates $m$ photons, we get :
\begin{equation}\label{eq:defG1}
    \hat{G}^{(1)} = -\sum_{m\neq0}\frac{\hat{H}_m}{m\omega_c},
\end{equation}
and similarly : 
\begin{equation*}
    \hat{V}_{X} = \sum_{m\neq0}\hat{H}_{m}.
\end{equation*}
So that finally Eq. \ref{eq:ExpansionWOrder1and2Deuxieme} allows to write using the argument that $\hat{H}_{m}$ creates $m$ photons and that the block diagonal term is the term which conserves the photon number: 
\begin{equation}\label{eq:VanVleckSecondOrderHamiltonian}
    \hat{W}^{(2)} = \sum_{m\neq0}\frac{\hat{H}_{m}\hat{H}_{-m}}{m\omega_c},
\end{equation}
where the simple change of variable $m\longmapsto -m$ has been operated.

Since the Floquet HFE is derived from the Van Vleck perturbation theory in Ref.\cite{eckardt_high-frequency_2015}, the equation verified by $\hat{G}^{(\ell)}$ will be identical to that of the kick operator in extended space. Thus, one is easily convinced that the equivalence between the Floquet HFE and cavity Van Vleck perturbation theory will extend to all order $\ell\in\N$.

\subsection{Details of the equivalence to the Quantum Floquet picture}\label{subsec:appQuantumFloquet}
Let us show that, for a single mode cavity, this formula is equivalent to the Floquet / Van Vleck one. To do so, we are going to link $\Hat{H}_m$ defined in \ref{eq:defFourierCoeffs} with the Quantum Floquet Hamiltonian $\hat{H}_{n,m}$. First, let us insert the closure relations $\sum_n \ket{n}\bra{n} = \mathbbm{1}$ in \ref{eq:defFourierCoeffs} : 
\begin{align*}
    \Hat{H}_m &\equiv  \frac{1}{T} \int_0^T \Hat{H}(t) e^{-\ii m\omega_c t}\ dt \\
    &=  \frac{1}{T} \int_0^T e^{\ii\omega_c t \hat{n}}\hat{H}e^{-\ii\omega_c t \hat{n}} e^{-\ii m\omega_c t}\ dt \\
    &= \frac{1}{T} \int_0^T \kern-0.5em \sum_{n,n^{\prime}\geq 0 } \kern-0.5em e^{\ii\omega_c t \hat{n}}\ket{n}\kern-0.25em\bra{n}\hat{H} \ket{{n^{\prime}}}\kern-0.25em\bra{{n^{\prime}}}e^{-\ii\omega_c t \hat{n}} e^{-\ii m\omega_c t}\ dt \\
    &= \sum_{n,n^{\prime}\geq 0}\hat{H}_{n,n^{\prime}} \underbrace{\frac{1}{T} \int_0^T e^{\ii (n-n^{\prime} - m ) \omega_c t}\ dt}_{= \delta_{n,n^{\prime}+m}}\ \otimes \ket{n}\bra{{n^{\prime}}} \\
    &= \left\{ \begin{array}{c}
         \sum_{n\geq 0 }\hat{H}_{n+m,n}\otimes \ket{n+m}\bra{n} \text{ if }m\geq0\\
         \sum_{n\geq 0 }\hat{H}_{n,n+|m|}\otimes \ket{n}\bra{n+|m|} \text{ if }m\leq0
    \end{array}\right.
\end{align*}
Thus,  at zeroth order in $\frac{1}{\omega}$ the equivalence of the Floquet and Brillouin Wigner HFE is clear, while for the first order the following calculations prove the equivalence :
\begin{align*}
    \sum_{\ell>0}\frac{\commut{\Hat{H}_{\ell}}{\Hat{H}_{-\ell}}}{\ell \omega} 
    &= \sum_{\ell>0}\kern-0.1em \sum_{n}\kern-0.1em \frac{1}{\ell \omega_c} \kern-0.2em \left( \hat{H}_{n+l,n}\hat{H}_{n,n+l}\kern-0.2em \otimes\kern-0.2em \ket{n \kern-0.2em + \kern-0.2em l}\kern-0.2em \bra{n\kern-0.2em +\kern-0.2em l} \right.\\
    &- \left. \hat{H}_{n,n+l}\hat{H}_{n+l,n}\otimes\ket{n}\bra{n}\right)  
\end{align*}
Performing the variable change $m = n+l$ in the previous sum, the first term corresponds to $n>m$ in \ref{eq:QuantumFloquetHamiltonian} while the second one corresponds to $m>n$. So that finally, when changing the names of the variables properly, one gets :
\begin{equation*}
    \sum_{\ell>0}\frac{\commut{\Hat{H}_{\ell}}{\Hat{H}_{-\ell}}}{\ell \omega_c} =  \sum_{{n}\neq{m}}\frac{ \hat{H}_{{n},{m}}  \hat{H}_{{m},{n}} }{({n}-{m}){\omega_c}}
\end{equation*}
Hence proving the equivalence, for a mono-mode cavity, of the Quantum Floquet effective Hamiltonian derived in Ref. \cite{li_effective_2022} and the effective Hamiltonian derived from Van Vleck perturbation theory, or equivalently from the Floquet HFE in the rotating frame.

Furthermore, in the multimode case, Eq.~\ref{eq:GcommutatorEquations} can be solved by projecting on the Fock basis of the light Hilbert space $\left|\mathbf{n}\right>$. The solution can be injected in Eq.~\ref{eq:ExpansionWOrder1and2Deuxieme} so as to recover Eq.~\ref{eq:QuantumFloquetHamiltonian} from Van Vleck perturbation theory.

\section{Details of the calculations for the SSH chain}\label{sec:AppendixSSH}

\subsection{Self-consistent equation for the effective hoppings of Eq.~\ref{eq:SSH_modelAtFirstOrderCompleteMeanField}}
The effective hoppings $\mathcal{V}_{eff}$ and $\mathcal{W}_{eff}$ appearing in the First order effective Hamiltonian within mean-field of Eq.~\ref{eq:SSH_modelAtFirstOrderCompleteMeanField} are defined self-consistently through :
\begin{widetext}
    \begin{subequations} \label{eq:SSH_effectiveHoppingsFirstOrderCompleteMF}
        \begin{align}
            \mathcal{V}_{eff} &= v_{eff} + \frac{2 v_{eff}^2 }{\omega_c} \left(f(\frac{g^2 b_0^2}{L}) \left<\hat{T}_{i.c.}\right> + f(-\frac{g^2 b_0^2}{L})  \left<\hat{T}_{i.c.}^{\dagger}\right> \right)\notag \\ &- \frac{2v_{eff} w_{eff} }{\omega_c} \left( f\parent{\frac{g^2}{L}b_0(1-b_0)}  \left<\hat{T}_{e.c.}^{\dagger}\right> + f\parent{-\frac{g^2}{L}b_0(1-b_0)}\left<\hat{T}_{e.c.}\right> \right), \\
            \mathcal{W}_{eff} &= w_{eff} - \frac{2 w_{eff}^2 }{\omega_c} \left(f(\frac{g^2 (1-b_0)^2}{L}) \left<\hat{T}_{e.c.}\right> + f(-\frac{g^2 (1-b_0)^2}{L})  \left<\hat{T}_{e.c.}^{\dagger}\right> \right)\notag \\ &+ \frac{2v_{eff} w_{eff} }{\omega_c} \left( f\parent{\frac{g^2}{L}b_0(1-b_0)}  \left<\hat{T}_{i.c.}^{\dagger}\right> + f\parent{-\frac{g^2}{L}b_0(1-b_0)}\left<\hat{T}_{i.c.}\right> \right).
        \end{align}
        \end{subequations}
\end{widetext}
The average $\left<\bullet\right>$ is taken on the ground state of the Hamiltonian defined in Eq.~\ref{eq:SSH_modelAtFirstOrderCompleteMeanField}. As such, the expectation values $\left<\hat{T}_{i.c./e.c.}\right> = \left<\hat{T}_{i.c./e.c.}\right>\left(\mathcal{V}_{eff}, \mathcal{W}_{eff}\right)$ depend on the effective hoppings. 

\subsection{Evaluation of the fluctuations of a one-body observable on a non-interacting ground state.}\label{app:CalculationFluctuationsOfJ}
We consider a non-interacting fermion problem which is diagonalized in the basis $\ket{\mu}$ with associated creation and annihilation operators $c_{\mu}^{\dagger}$ and $c_{\mu}$. We consider a one body operator : 
\begin{align*} 
\hat{J} &= \sum_{\mu,\nu}  J_{\mu,\nu} c_{\mu}^{\dagger}c_{\nu},
\end{align*}
and we aim to compute its fluctuations in the ground state of the system : ${\Delta J }^2 \equiv \langle \hat{J}^2\rangle -\langle \hat{J}\rangle^2  $. Let us first introduce the decomposition of $\hat{J}$ into its diagonal and off-diagonal part : 
\begin{align*} 
    \hat{J} &= \hat{J}_{D}+ \hat{J}_{X},
\end{align*}
where $\hat{J}_{D} \equiv \sum_{\mu}J_{\mu,\mu} \hat{n}_{\mu}$, and thus by definition $\hat{J}_{X} \equiv \sum_{\mu \neq \nu}  J_{\mu,\nu} c_{\mu}^{\dagger}c_{\nu} $. Clearly, the ground state $\ket{\psi_0}$ is an eigenstate of $\hat{J}_{D} $ and the average value of $\hat{J} $ is given by :
\begin{equation} \label{eq:averageValuJandJD}
\left< \hat{J}  \right>  = \left< \hat{J}_{D}  \right>.
\end{equation}
Then, the average value of the square reads :
\begin{align*} 
    \left< \hat{J}^2  \right> &= \left< \left(\hat{J}_{D} +\hat{J}_{X} \right)^2  \right>\\
    &= \left<\hat{J}_{D}^2  \right> + \left<\hat{J}_{X}^2  \right> + \left< \left\{\hat{J}_{D} , \hat{J}_{X} \right\}  \right>\\
    &= \left<\hat{J}_{D}  \right>^2 + \left<\hat{J}_{X}^2  \right> +2 \cancel{\left< \hat{J}_{X}\right>} \left<  \hat{J}_{D}   \right>\\
    &= \left<\hat{J}  \right>^2 + \left<\hat{J}_{X}^2  \right> ,
\end{align*}
where it has been used that $\ket{\psi_0}$ is an eigenstate of $\hat{J}_{D} $ to go from the 3rd to the 4th line. From this, the fluctuations of $J$ are given by : 
\begin{align*} 
    {\Delta J }^2 &= \left<\hat{J}_{X}^2  \right>\\
    &= \sum_{\mu \neq \nu}\sum_{\rho \neq \eta}  J_{\mu,\nu} J_{\rho,\eta} \left<c_{\mu}^{\dagger}c_{\nu}    c_{\rho}^{\dagger}c_{\eta} \right> 
\end{align*}
then, if $\rho = \nu$ and $\eta = \mu$ are not verified, the expectation value above vanishes as it does not conserve the quantum numbers $n_{\mu}$.
\begin{align*} 
    {\Delta J }^2  &= \sum_{\mu \neq \nu}\sum_{\rho \neq \eta}  J_{\mu,\nu} J_{\rho,\eta} \left<c_{\mu}^{\dagger}c_{\nu}    c_{\rho}^{\dagger}c_{\eta} \right>\\
    &= \sum_{\mu \neq \nu} J_{\mu,\nu} J_{\nu,\mu} \left<c_{\mu}^{\dagger}c_{\nu}    c_{\nu}^{\dagger}c_{\mu} \right> \\
    &= \sum_{\mu \neq \nu} J_{\mu,\nu} J_{\nu,\mu} \left<c_{\mu}^{\dagger}c_{\mu}c_{\nu}    c_{\nu}^{\dagger} \right> \\
    &= \sum_{\mu \neq \nu} J_{\mu,\nu} J_{\nu,\mu} \left<c_{\mu}^{\dagger}c_{\mu}(1  -  c_{\nu}^{\dagger}c_{\nu}) \right> \\
    &= \sum_{\mu \neq \nu} J_{\mu,\nu} J_{\nu,\mu} \left<\hat{n}_{\mu}(1  -  \hat{n}_{\nu}) \right> ,
\end{align*}
so that finally : 
\begin{equation} \label{eq:expressionDeltaJ}
    {\Delta J }^2  = \sum_{\mu \neq \nu} J_{\mu,\nu} J_{\nu,\mu} \left<\hat{n}_{\mu}\right>(1  -\left<  \hat{n}_{\nu} \right> ) ,
\end{equation}
since the ground state is an eigenstate of the occupation operators.

\subsection{Calculations for the entropy}
To compute the Renyi entenglement entropy in the ground state of the SSH chain coupled to an off-resonant cavity, we use the result obtained in Sect.~\ref{sec:VanVleck}. In the particular case of the SSH model, the $\hat{\mathcal{J}}_n$ operator defined in Eq.~\ref{eq:Nphoton_PseudoCurrent} is given by : 
% \begin{widetext}
\begin{align} 
    \hat{\mathcal{J}}_n &= \frac{\ii^n}{n\sqrt{n!}}\left[\frac{v}{\omega_c} e^{-\frac{g^2b_0^2}{2L}} \left( \frac{gb_0}{\sqrt{L}} \right)^n \hat{T}_{i.c.} \right. \notag \\
    &-\left. \frac{w}{\omega_c} e^{-\frac{g^2(1 - b_0)^2}{2L}} \left( -\frac{g(1-b_0)}{\sqrt{L}} \right)^n \hat{T}_{e.c.}\right] + h.c. \label{eq:SSH_Jn}
\end{align}
% \end{widetext}

In particular, it is an hermitian operator. Then, one defines the complex variables : 
\begin{align} \label{eq:Def_smallJn}
j_n(k) &= \ii^n\left[\frac{v}{\omega_c} e^{-\frac{g^2b_0^2}{2L}} \left( \frac{gb_0}{\sqrt{L}} \right)^n\right.\notag \\
& -\left. \frac{w}{\omega_c} e^{-\frac{g^2(1 - b_0)^2}{2L}} \left( -\frac{g(1-b_0)}{\sqrt{L}} \right)^n e^{\ii k }\right],
\end{align}
so that, in spinor form, the operator $\hat{\mathcal{J}}_n$ reads : 
\begin{align*}  \label{eq:CurrentInSpinorForm}
    \hat{\mathcal{J}}_n &= \frac{1}{n\sqrt{n!}}  \sum_k \left( \begin{array}{cc} c_{k,A}^{\dagger}  & c_{k,B}^{\dagger}  \end{array} \right) \left( \begin{array}{cc} 0 & j_n(k)   \\  j_n(k)^{\ast} & 0   \end{array} \right)\left( \begin{array}{c} c_{k,A} \\ c_{k,B}  \end{array} \right) 
    % = \frac{1}{\sqrt{n!}}  &\sum_k \left( \begin{array}{cc} c_{k,A}^{\dagger}  & c_{k,B}^{\dagger}  \end{array} \right) \left( \text{Re}j_n(k) \tau_x - \text{Im}j_n(k) \tau_y  \right)\left( \begin{array}{c} c_{k,A} \\ c_{k,B}  \end{array} \right) 
\end{align*}
The effective SSH hamiltonian of Eq.~\ref{eq:SSH_modelAtFirstOrderCompleteMeanField} is diagonalized by introducing the creation and destruction operators : 
\begin{equation*}
    \left( \begin{array}{c} d_{k,+}  \\   d_{k,-} \end{array} \right) \equiv e^{\ii\frac{\pi}{4}\tau_{y}}e^{\ii\frac{\theta_k}{2}\tau_{z}} \left( \begin{array}{c} c_{k,A} \\ c_{k,B}  \end{array} \right) ,
\end{equation*}
where $\theta_k \equiv \text{arg}\parent{\mathcal{V}_{eff}  - \mathcal{W}_{eff} e^{-\ii k} }$ and $\tau_{x,y,z}$ are the Pauli matrices. One finds the expression of $\hat{\mathcal{J}}_n $ :
% \begin{widetext}
    \begin{align} 
        \hat{\mathcal{J}}_n \kern-0.1em &= \kern-0.1em \frac{1}{n\sqrt{n!}}\sum_{k} \kern-0.1em \left|j_n(k)\right|\kern-0.1em  \left( \kern-0.25em\begin{array}{cc} d_{k,+}^{\dagger}  &   d_{k,-}^{\dagger} \end{array} \kern-0.25em\right)\kern-0.1em \left[ \cos\left(\text{arg}j_n(k)\kern-0.1em - \kern-0.1em\theta_k\right)\kern-0.1em \tau_z\right.\notag \\ & + \left.\sin{\left(\text{arg}j_n(k) - \theta_k\right) \tau_y }  \right]\left( \begin{array}{c} d_{k,+}  \\   d_{k,-} \end{array} \right) . \label{eq:spinorExpressionForJ_n}
    \end{align}
% \end{widetext}
    Then, using the result of App.~\ref{app:CalculationFluctuationsOfJ} and in particular that of Eq.~\ref{eq:expressionDeltaJ}, the fluctuations of $\hat{\mathcal{J}}_n$ are given by : 
\begin{equation} \label{eq:formulaFluctuationsOfJ_n}
\left<\hat{\mathcal{J}}_n^2\right> - \left<\hat{\mathcal{J}}_n\right>^2 = \frac{1}{n! n^2}\sum_{k}  \left|j_n(k)\right|^2 \sin^2{\left(\text{arg}j_n(k) - \theta_k\right)  } ,
\end{equation}
which is easily evaluated numerically.

\subsection{Details of the time-dependent mean-field}\label{app:timeDependentMF} % for Sect. \ref{sec:drivenSSH_Rabi}
This appendix details the procedure used to compute the dynamics of electron observables in Sect. \ref{sec:drivenSSH_Rabi} through a time-dependent mean-field.   

In the $n$ photon sector, the dynamics of the electrons are controlled by the effective Hamiltonian :
\begin{equation} \label{eq:initialHamiltonian_AppMF}
\hat{H}_{eff,n}^{[1]} = \sum_{i,j} t_{i,j}^{n} c_i^{\dagger}c_j + \sum_{ijkl} V_{ijkl}^{n} c_i^{\dagger}c_jc_k^{\dagger}c_l,
\end{equation}
where $ t_{i,j}^{n}$ and $V_{ijkl}^{n}$ are coefficients obtained from the Floquet HFE and whose analytical expression, while cumbersome, can easily be evaluated numerically. 

The effective Floquet Hamiltonian for the n-photon sector $\hat{H}_{F,n} \equiv \left<n\right|\hat{H}_{F}\left|n\right>$ is given by Eq.~\ref{eq:NphotonSector_FirstOrderHamiltonian}. The dynamics inside each photon sector will be treated using a time-dependent mean-field decoupling of the cavity mediated interaction. Under this approximation, the evolution operator in the $n$-photon sector $\hat{U}_n(t) \equiv e^{-\ii \hat{H}_{F,n} t}$, is approximated as :
\begin{equation} \label{eq:TimeDependentMeanField_Evolution}
    \hat{U}_n(t) \simeq \mathcal{T}e^{-\ii\int _0^t \hat{H}_{eff,n}^{MF}(t^{\prime}) dt^{\prime} }
\end{equation} 
where the mean-field decoupled Hamiltonian is obtained, within the same logic as in Sect. \ref{sec:SSH_equilibrium}, through the Hatree-Fock decoupling of the hamiltonian of Eq.~\ref{eq:initialHamiltonian_AppMF} to obtain :  
\begin{align} \label{eq:TimeDependentMeanFieldHamiltonian}
    \hat{H}_{eff,n}^{MF}(t)  &=  \sum_{i,j} t_{i,j}^{n} c_i^{\dagger}c_j + \sum_{ijkl} V_{ijkl}^{n} \left( \left<c_i^{\dagger}c_j\right>_{\kern-0.2em n} \kern-0.5em (t) c_k^{\dagger}c_l \right.  \\ & \kern-3em\left.  +  \left< c_k^{\dagger}c_l\right>_{\kern-0.2em n} \kern-0.5em (t) c_i^{\dagger}c_j   + \left<c_i^{\dagger}c_l\right>_{\kern-0.2em n} \kern-0.5em (t) c_jc_k^{\dagger}+ \left<c_jc_k^{\dagger}\right>_{\kern-0.2em n} \kern-0.5em (t) c_i^{\dagger}c_l \right),\notag
\end{align}
where a scalar term has been discarded and $\left< \bullet \right>_{\kern-0.1em n} \kern-0.25em (t)$ designates the average on the initial state evolved through Eq.~\ref{eq:TimeDependentMeanField_Evolution}. 

The dynamics generated by Eq.~\ref{eq:TimeDependentMeanField_Evolution} conserve the expectation value of the energy, which in the Hartree-Fock treatement is given by : 
\begin{align} 
    \left<\hat{H}_{eff,n}\right>_{\kern-0.2em n} \kern-0.5em (t) &=  \kern-0.25em \sum_{i,j}\kern-0.2em t_{i,j}^{n} c_i^{\dagger}c_j \kern-0.2em+ \sum_{ijkl} V_{ijkl}^{n} \left( \left<c_i^{\dagger}c_j\right>_{\kern-0.2em n} \kern-0.5em (t) \left<c_k^{\dagger}c_l\right>_{\kern-0.2em n}\kern-0.5em (t)\right.\notag\\&+ \left. \left<c_jc_k^{\dagger}\right>_{\kern-0.2em n} \kern-0.5em (t) \left<c_i^{\dagger}c_l\right>_{\kern-0.2em n}\kern-0.5em (t) \right)\label{eq:HartreefFockEnergy}.
\end{align}
Indeed, taking the derivative with respect to time of the previous expression and using Eq.~\ref{eq:TimeDependentMeanField_Evolution} to show that :
\begin{align*} 
\frac{d}{dt}\left<c_i^{\dagger}c_j\right>_{\kern-0.2em n} \kern-0.5em (t) &= \ii \left< \left[\hat{H}_{eff,n}^{MF}(t), c_i^{\dagger}c_j\right] \right>_{\kern-0.2em n}\kern-0.5em (t).
\end{align*}
All the terms generated by the differentiation of Eq.~\ref{eq:HartreefFockEnergy} lead to : 
\begin{align*} 
    \frac{d}{dt}\left<\hat{H}_{eff,n}\right>_{\kern-0.2em n} \kern-0.5em (t) &= \ii \left< \left[\hat{H}_{eff,n}^{MF}(t), \hat{H}_{eff,n}^{MF}(t)\right] \right>_{\kern-0.2em n}\kern-0.5em (t)\\
    &=0.
\end{align*}
This conservation serves as a witness of the precision of the numerical integration of the dynamics. Furthermore, unlike in Sect.~\ref{sec:SSH_equilibrium} no channels of the Hartreef-Fock decoupling are discarded so as to conserve the energy throughout the dynamics.

For any one-body operator on the fermions $\hat{O}$, a $2L\times2L$ matrix $\mathbf{O}$ can be defined such that $\hat{O} = \sum_{i,j}  (\mathbf{O})_{i,j}c_{i}^{\dagger}c_{j} $. In particular, the $2L\times2L$ matrix whose matrix elements are given by $\left< c_{i}^{\dagger}c_{j}\right>_n(t)$ is denoted $\mathbf{C_n}(t)$ . One easily shows that the expectation value of a one-body operator $\hat{O}$ described by the matrix $\mathbf{O}$ and evolved in the n-photon sector can be computed as :
\begin{equation} \label{eq:AverageOneBodyOperator}
\left<\hat{O}\right>_{\kern-0.2em n} \kern-0.5em (t) = \text{tr} \left( \mathbf{O}^{T} \mathbf{C_n}(t)\right),
\end{equation}
where the trace is taken on the $2L$-dimensionnal space underlying the matrices and $\mathbf{O}^{T}$ refers to the transpose of the matrix $\mathbf{O}$ . In particular, $\hat{H}_{eff,n}^{MF}$ can be computed from Eq.~\ref{eq:TimeDependentMeanFieldHamiltonian} at each step. 

From this, we know that in the time-dependent mean-field approximation only the dynamics of $\mathbf{C_n}$ need to be computed. Using Eq.~\ref{eq:CommutaionRelatioHoppingFermions}, one shows that the equation of motion for $\mathbf{C_n}(t)$ is given by :
\begin{equation*}% \label{eq:TimeDependentMeanField_EquationOfMotionC}
\forall i,j, \ (\frac{d}{dt}\mathbf{C_n})_{i,j} = - \ii \left(\left[ (\mathbf{H_n}^{MF})^{T}(t), \mathbf{C_n}(t)  \right]\right)_{i,j}.
\end{equation*}
The differential equation is initialized such that at $t = 0$ the matrix $\mathbf{C_n}(t=0)$ be given by the equilibrium expectation value of $\left< c_{i}^{\dagger}c_{j}\right>$ which are computed from the diagonalization Open Boundary Condition SSH hamiltonian obtained from the mean-field value of the coefficients $\mathcal{V}$ and $\mathcal{W}$.

\section{On the link between the single-electron and many-body problem}\label{app:singleElectron}
In Refs. \cite{perez-gonzalez_light-matter_2025, bomantara_quantum-vacuum-protected_2025}, %\cite{perez_gonzalez_quantum_2024}
starting from a representation of the light-matter Hamiltonian as : 
\begin{align*} 
\hat{H} &= \sum_{im,jn} H_{im,jn} \left|m\right>\left<n\right|\otimes c_i^{\dagger}c_j,
\end{align*}
the matrix $\mathbf{H} = H_{im,jn}$ is diagonalized numerically or analytically, so that one has a unitary $\mathbf{U}$ so that $\mathbf{U}\mathbf{H}\mathbf{U}^{\dagger} = \mathbf{D}$ with $D_{im,jn} = D_{im} \delta_{ij}\delta_{mn}$. As such, the Hamiltonian can be written : 
\begin{align*} 
\hat{H} &= \sum_{im,jn}\left(\mathbf{U}^{\dagger}\mathbf{H}\mathbf{U}\right)_{im,jn}  \left|m\right>\left<n\right|\otimes c_i^{\dagger}c_j\\
&= \sum_{kl}D_{kl} \sum_{im,jn} U_{kl,im}^{\ast} U_{kl,jn}\left|m\right>\left<n\right|\otimes c_i^{\dagger}c_j.
\end{align*}
And then, denoting $\hat{P}$ the projection onto the $1$ electron subspace, which acts on hoppings as $\hat{P}c_i^{\dagger}c_j\hat{P} = \left|i\right>\left<j\right|$. So that the Hamiltonian on the single electron subspace reads :
\begin{align*} 
\hat{P}\hat{H}\hat{P} &= \sum_{kl}D_{kl} \sum_{im,jn} U_{kl,im}^{\ast} U_{kl,jn}\left|m\right>\left<n\right|\otimes\left|i\right>\left<j\right| \\
&= \sum_{kl}D_{kl} \sum_{im,jn} U_{kl,im}^{\ast} U_{kl,jn}\left( \left|m\right>\otimes \left|i\right>  \right) \left(\left<n\right|\otimes\left<j\right|  \right)  \\
&= \kern-0.1em \sum_{kl}D_{kl}\kern-0.1em \left(\kern-0.1em \sum_{im} U_{kl,im}^{\ast} \left|m\right>\otimes \left|i\right> \kern-0.1em \right) \kern-0.3em \left(  \vphantom{\sum_{im} U_{kl,im}^{\ast} \left|m\right>\otimes \left|i\right>} \right. \kern-0.3em
    \sum_{jn} U_{kl,jn} \left<n\right|\otimes\left<j\right|  \kern-0.3em \left. \vphantom{\sum_{im} U_{kl,im}^{\ast}  \left|m\right>\otimes \left|i\right> } \right)  ,
\end{align*}
and $\left\{\sum_{im} U_{kl,im}^{\ast} \left|m\right>\otimes \left|i\right>  \right\}_{kl}$ thus forms a basis of diagonalization of $\hat{P}\hat{H}\hat{P}$. Howver, outside the $1$ electron subspace  $\left|m\right>\left<n\right|\otimes c_i^{\dagger}c_j$ is not $\left(\left|m\right>\otimes c_i^{\dagger}\right) \left(\left<n\right|c_j\right)$ (which has no meaning). Thus, the many-body Hamiltonian is not of the form :
\begin{align*} 
\hat{H} &\neq \sum_{kl} D_{kl} \left|l\right>\left<l\right| \otimes c_k^{\dagger}c_k.
\end{align*}

Usually, the correspondence between single-particle physics and many-body physics of non-interacting fermions is justified by constructing a unitary transformation on the full Hilbert space from the one defined on the single particle subspace. Indeed, if the hamiltonian is of the form :
\begin{align*} 
\hat{H} &= \sum_{ij} H_{i,j} c_i^{\dagger}c_j
\end{align*} 
and the matrix $\mathbf{H}$ is diagonalized by the unitary transformation $\mathbf{U} = e^{\ii\mathbf{G}}$ where $\mathbf{G}$ is an hermitian matrix. One defines the diagonalization unitary $\hat{U}$ on the full Hilbert space through :
\begin{align*} 
\hat{U} &\equiv e^{\ii \sum_{i,j} G_{i,j} c_i^{\dagger}c_j}\\
&= e^{\ii \hat{G}},
\end{align*}
so that the Hamiltonian is transformed under this unitary transformation as : 
\begin{align*} 
\hat{U}\hat{H}\hat{U}^{\dagger} &= e^{\ii \text{ad}_{\hat{G}}} \hat{H}.
\end{align*}
Then, using Eq.~\ref{eq:CommutaionRelatioHoppingFermions}, one shows that if $\hat{A} = \sum_{i,j} A_{i,j}c_i^{\dagger}c_j$ and $\hat{B} = \sum_{i,j} B_{i,j}c_i^{\dagger}c_j$, then their commutator reads : 
\begin{align} 
\left[ \hat{A},\hat{B}  \right] &= \sum_{i,j}  \left(\left[\mathbf{A},\mathbf{B}\right]  \right)_{i,j} c_i^{\dagger}c_j,\label{eq:OneBodyCommutators}
\end{align}
thus :
\begin{align*} 
    \hat{U}\hat{H}\hat{U}^{\dagger} &= \sum_{i,j}\left(e^{\ii \text{ad}_{\mathbf{G}}} \mathbf{H}\right)_{i,j}c_i^{\dagger}c_j \\
    &=  \sum_{i,j}\left( \mathbf{U} \mathbf{H}\mathbf{U}^{\dagger}\right)_{i,j}c_i^{\dagger}c_j
\end{align*}
is diagonal. This why single particle physics is enough to describe not interacting fermionic system. However, Eq.~\ref{eq:OneBodyCommutators} breaks down for $n\geq2$-body operators and the correspondence exploited above is no more.

Trying to perform a similar procedure, by promoting the unitary matrix $\mathbf{U} = U_{im,jn}$ to a unitary transformation on the full light-matter Hilbert space by defining :
\begin{align*} 
\hat{U} &= \exp\left(\ii \sum_{im,jn}  G_{im,jn} \left|m\right>\left<n\right|\otimes c_i^{\dagger}c_j\right),
\end{align*}
so that the action of the unitary transformation on the Hamiltonian is :
\begin{align*} 
    \hat{U}\hat{H}\hat{U}^{\dagger} &= e^{\ii \text{ad}_{\hat{G}}} \hat{H}\\
    &= \hat{H} + \ii \left[\hat{G},\hat{H} \right] + \dots 
\end{align*}
And, using once again Eq.~\ref{eq:commutatorIdentityTensorProduct} to compute the commutator of a tensor product of operators, one finally finds that the first order correction to the Hamiltonian is :
\begin{align*} \label{eq:commutatorGandH}
    \left[\hat{G},\hat{H} \right]\kern-0.25em &= \kern-0.25em\frac{1}{2}\kern-0.2em\sum_{im,jn}\kern-0.4em \left( \left[\mathbf{G}, \mathbf{H}\right] + \left[\mathbf{G}^{TL}, \mathbf{H}^{TL}\right]  \right)_{im,jn} \left|m\right>\kern-0.2em\left<n\right|\kern-0.2em\otimes\kern-0.2em c_i^{\dagger}c_j\notag \\
    &+   \kern-0.75em \sum_{\substack{im,jn  \\ i^{\prime}\kern-0.1em m^{\prime} \kern-0.2em,j^{\prime}\kern-0.1em n^{\prime} }  } \kern-0.75em \left(G_{im,jn^{\prime}} H_{i^{\prime}n^{\prime},j^{\prime}n} - G_{in^{\prime},jn}H_{i^{\prime}m,j^{\prime}n^{\prime}} \right)\left|m\right>\kern-0.2em\left<n\right| \\
    &\otimes  c_i^{\dagger}c_jc_{i^{\prime}}^{\dagger}c_{j^{\prime}},
\end{align*}
where the transpose with respect to the light indices has been defined $\left(\mathbf{H}^{TL} \right)_{im,jn} = H_{in,jm}$. From this, equation we can see that any treatment of the problem which only considers the single-electron subspace discards all the interactions mediated by the cavity. In the case of the HFE, it is equivalent to truncating the HFE at order zero, which has been shown in Sect. \ref{sec:SSH} to not be sufficient to describe the physics of the system at high light matter coupling.

%\bibliography{references.bib}
%\bibliographystyle{apsrev4-1}

%

\end{document}